\documentclass[10pt, final, journal, letterpaper, twocolumn]{IEEEtran}
\usepackage{times}
\usepackage{cite}
\usepackage{amsmath}
\usepackage{array}
\usepackage{amssymb}

\usepackage{stfloats}
\usepackage{rotating,threeparttable,booktabs}
\usepackage{bm}
\usepackage{dcolumn,booktabs}
\usepackage{multirow}
\usepackage{graphicx}
\usepackage{color}
\usepackage{algorithm}
\usepackage{algorithmic}
\usepackage{booktabs}
\usepackage{multirow}
\usepackage{makecell}
\usepackage{graphicx}
\usepackage{epstopdf}

\ifCLASSOPTIONcompsoc
\usepackage[caption=false,font=normalsize,labelfont=sf,textfont=sf]{subfig}
\else
\usepackage[caption=false,font=footnotesize]{subfig}
\fi
\newcommand{\RNum}[1]{\uppercase\expandafter{\romannumeral #1\relax}}
\ifCLASSINFOpdf
\else

\fi

\usepackage{amsmath,amsthm,amssymb,amsfonts}
\makeatletter
\thm@headfont{\sc}
\makeatother

\usepackage{url}
\begin{document}

\title{Deep Reinforcement Learning Based Dynamic Power and Beamforming Design for Time-Varying Wireless Downlink Interference Channel}
\author{\authorblockA {Mengfan Liu, Rui Wang, \textit{Senior Member IEEE}}
\thanks{M. Liu and R. Wang are with the Department of Information and Communications at Tongji University, Shanghai, 201804, China. R. Wang is also with the Shanghai Institute of Intelligent Science and Technology, Tongji University, Shanghai, China. R. Wang is the \emph{corresponding author}, (e-mails: liumengfantj@126.com, ruiwang@tongji.edu.cn). }}
\maketitle

\begin{abstract}
With the high development of wireless communication techniques, it is widely used in various fields for convenient and efficient data transmission. Different from commonly used assumption of the time-invariant wireless channel, 
we focus on the research on the time-varying wireless downlink channel to get close to the practical situation. Our objective is to gain the maximum value of sum rate in the time-varying channel under the some constraints about cut-off signal-to-interference and noise ratio (SINR), transmitted power and beamforming. In order to adapt the rapid changing channel, we abandon the frequently used algorithm convex optimization and  deep reinforcement learning algorithms are used in this paper. From the view of the ordinary measures such as power control, interference incoordination and beamforming, continuous changes of measures should be put into consideration while sparse reward problem due to the abortion of episodes as an important bottleneck should not be ignored. Therefore, with the analysis of relevant algorithms, we proposed two algorithms, Deep Deterministic Policy Gradient algorithm (DDPG) and hierarchical DDPG, in our work. As for these two algorithms, in order to solve the discrete output, DDPG is established by combining the Actor-Critic algorithm with Deep Q-learning (DQN), so that it can output the continuous actions without sacrificing the existed advantages brought by DQN and also can improve the performance. Also, to address the challenge of sparse reward, we take advantage of meta policy from the idea of hierarchical theory to divide one agent in DDPG into one meta-controller and one controller as hierarchical DDPG. Our simulation results demonstrate that the proposed DDPG and hierarchical DDPG performs well from the views of coverage, convergence and sum rate performance.
\end{abstract}


\IEEEpeerreviewmaketitle

\section{Introduction}
With the rapid development of wireless communication technology, it is challenging to improve the performance of system on the basis of keeping high resource utilization efficiency. Cellular communication, as a representative wireless communication network, can organize the base stations (BSs) rationally, but it is still inevitable to receive the noise due to the reflection, scattering or refraction of electromagnetic waves in the propagation and also the interference from other base stations so that the performance is limited. Therefore, it deserves to be the priority to deal with interference and noise in the wireless communication field. At present, power allocation\cite{ wong1999multiuser} and beamforming design\cite{ cox1987robust } with convex optimization perform well to solve this problem in previous works\cite{ liu2019energy, xiu20201reconfigurable, du2020wirelessly}. However, convex optimization method\cite{ boyd2004convex } requires a huge amount of computational power so it is necessary to simplify the channel into the time invariant one or time-varying channel with fixed routes\cite{4,5,6,7}, which violates the actual situation. In most actual communication scenarios, it is impossible to keep users from moving, and because of the slightly rapid movement of the objects, the coherent time of the channel is reduced with the growth of doppler spread\cite{2}, which leads that the duration of the signal is longer than the coherent time of the channel, the channel impulse response of each transmission path changes rapidly with time, and wireless channel gradually evolves into a time-frequency dual selective fading channel, namely time-varying channel\cite{3}. And It is exactly what we concerned about in this paper. In this way, it is important to propose an algorithm in order to make the approximately real-time interaction with the environment easy. So we focus on the deep reinforcement learning (DRL) and consider the problem appearing consequently, such as how to output the continuous actions for this model and how to improve sparse reward problem in the training process.

Therefore, in order to make the solutions closer to the practical situation, we investigate the time-varying wireless downlink channel \cite{8,9,10,11,12} in this paper. Here power control (PC), interference coordination (IC) and beamforming (BF) are still the means to coordinate the interference with an aim to improve the network performance. However, it is noted that the time-varying channel introduces some new constraints on the design problem and makes the problem non-convex and hard to solve. To address the encountered challenges, we propose a new deep reinforcement learning approach to ensure the output of continuous actions, and try to combine it with the hierarchical theory to solve the long-timescale sparse reward problems. Basically, these two proposed methods can adapt the rapid changes of the channels and also bring the better performances than the present industry standard and some traditional deep reinforcement learning techniques.


\subsection{Prior Works}
There are many works about the non-convex problem solved by value-based reinforcement learning. In \cite{ ye2019power}, dynamic programming method, one of solutions to the model-based problems, was applied to gain the offline power policy. As for the model-free problems in which the probability of state transmission is unknown, \cite{ ye2019power} researched about the application of SARSA algorithm to implement the online power policy and \cite{ liao2020dynamic } tried to improve the dynamic spectrum access to select the channel based SARSA. The optimization in the downlink was studied in \cite{16}, applying the tabular reinforcement learning. And \cite{17} discussed the joint beamforming and interference coordination as the measures rather than the independent use of these two methods.

On the other hand, policy-based reinforcement learning plays an important role in the optimization of wireless channel. \cite{ chai2019joint } focused on the joint rate and power optimization in the fading channel by using policy gradient with deterministic policy. Innovatively, Actor-Critic method, as the combination of policy gradient and tabular Q-learning, performed well in the problem of user scheduling and resource allocation in \cite{ wei2017user }.

In the meanwhile, the introduction of deep neural network produces a revolutionary change. In \cite{13,14,15}, the performances in the downlink was the focus, on which power control and beamforming respectively were studied by simply using deep neural networks. Furthermore, \cite{18} proposed joint power control, interference coordination and beamforming as measures in downlink by using deep Q-learning (DQN). It should be noted that deep Q-learning is one kind of off-policy temporal difference algorithm with deep neural network but it is limited by the discrete output.

In our paper, different from previous studies, we research about deep deterministic policy gradient algorithm (DDPG)\cite{28} with the ideas of DQN and Actor-Critic, making continuous outputs possible. Also, the hierarchical theory\cite{29,30,31,32,33} is a the-state-of-art research direction for the solution to the long-timescale sparse reward problem. Meta policy was applied in \cite{34} to combine with DQN and improve the performance. Therefore, we try to employ the DDPG and hierarchical DDPG (h-DDPG) in the time-varying wireless downlink interference channel for the joint online optimization of the power control, interference coordination and beamforming design.

\subsection{Contribution}
With the applications of two proposed deep reinforcement learning techniques in the time-varying wireless downlink channel, the novel contributions made by this paper are as follows:
\begin{itemize}
	\item 
Due to the online learning characteristic of DRL, the channel information acquirement is not required for the joint optimization of the power control, interference coordination and beamforming design. This alleviates a huge amount of information exchange overhead and makes the adaptation of time-varying environment possible. In addition, when multiple base stations are involved, the complexity of previous algorithms like industry standard algorithm is daunting, which can be improved due to the introduction of neural networks.
	\item Our proposed DRL is able to output the continuous actions of joint power control and interference coordination, which breaks through the bottleneck that the traditional algorithms can only produce discrete measures. The continuous output of PC and IC are more flexible and also closer to the real situations. Furthermore, the performance of the model is improved due to the more maneuverable adjustment techniques.
	\item We solve the sparse reward problem since there are not enough positive rewards as feedback due to the abortion of the unsuccessful episodes to some degree by applying the hierarchical DDPG. The introduced hierarchical theory reduces the influence of the sparse reward problems and improve the performance more than simple DRL algorithms.
\end{itemize}

The rest of the paper is organized as follows. The system model and problem formulation is described in Section \ref{Sytem_mdoel} while the basic knowledge of deep reinforcement learning is in Section \ref{Preliminary}. Section \ref{DDPG} describes concretely DDPG based on the time-varying channel while in Section \ref{Hiera_DDPG} we introduce the hierarchical DDPG in the practical application. At last, the simulation and results is in Section \ref{Simulations} and the conclusion is presented in Section \ref{conclusions}.

\emph{Notations}: 
$E$ denotes the statistical expectation and ${\bf F}$ is the generic matrix. For any generic column vector ${\bf h}$, ${\bf h}_{i, j}$ denotes the vector from the $i$-th start point to the $j$-th end point, and ${\bf h}^T$ is the transpose of ${\bf h}$. ${\bf h}[t]$ represents the vector at the time step $t$. For any generic scalar $x$, $|x|$ denotes the absolute value of $x$ and $x_i$ is the $i$-th value depending on the specific object. $\mathcal{CN}\left( {0,{\sigma _n}^2} \right)$ represents the normal distribution with zero-mean and variance ${{\sigma _n}^2}$.

\section{System Model and Problem Formulation}\label{Sytem_mdoel}

\begin{figure*}[!t]
	\centering 
	\includegraphics[scale=0.40]{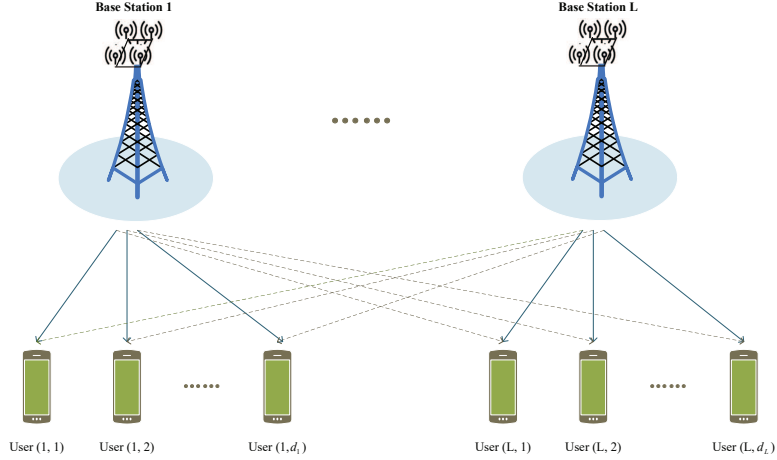} 
	\caption{The considered time-varying multi-cell downlink interference channel.}
	\label{Fig.1}
\end{figure*}

In this section, we present the system model and the formulation of problem we concern in this paper. To be brief, the basic setup of transmit powers and beamforming while the organization of base stations is introduced, following by the detailed description of constraints and the goal of our problem.

\subsection{System Model}

In the network model as illustrated in Fig. \ref{Fig.1}, there are $L$ BSs with $L \geq 2$,  
and the $l$-th base station serves totally ${d_l}$ users with  ${d_l} \geq 1$. Here it should be noted that the user equipment (UE), scattering randomly at the serving area, cannot be served by more than one serving base station. As for the serving area, cellular network\cite{35} divides a large area into several small cells in which there is one base station at least. In our model shown in Fig.\ref{Fig.1}, we simplify the cell to a circle with the radius $R/2$ and it is allowable to overlap the cells, which makes limit frequency resources can be reused in a range.

When it comes to each base station, the model adopts the multi-antenna setup and there is a uniform linear array of $M$ antennas. In the meanwhile, the user equipment employs single antenna.
Therefore, for the $i$-th user equipment from all users served by the $l$-th BS, the received signal at the time step $t$, denoted by ${y_{\left(l,i\right)}[t]}$, consists of desired received signal from serving BS, the sum of received interference and the random noise. The formula can be expressed as
\begin{equation}
\begin{aligned}
{y_{\left(l,i\right)}[t]} =&{\bf h}_{l,\left(l,i\right)}^T[t]{{\bf f}_{l,\left(l,i\right)}}[t]{x_{l,\left(l,i\right)}}[t]\\
 &+ \sum\limits_{l \ne b} {{\bf h}_{b,\left(l,i\right)}^T[t]{{\bf f}_{b,\left(l,i\right)}}[t]{x_{b,\left(l,i\right)}}[t]}
+ {n_{\left(l,i\right)}}[t]
\end{aligned}
\end{equation}
where ${{\bf h}_{b,\left(l,i\right)}[t]} \in {\mathbb{C}^{M \times 1}}$ denotes the channel vector between the $i$-th user served by the $l$-th BS and the $b$-th BS at the time step $t$, ${x_{b,\left(l,i\right)}[t]}$ denotes the transmitted signals at the time step $t$ from the $b$-th BS to the $i$-th user served by $l$-th BS, satisfying $E\left[ {|{x_{b,\left(l,i\right)}[t]}{|^2}} \right] = P_{b,\left(l,i\right)}[t]$ with $P_{b,\left(l,i\right)}[t]$ being the transmitted power associated with ${x_{b,\left(l,i\right)}[t]}$,
${{\bf f}_{b,\left(l,i\right)}[t]} \in {\mathbb{C}^{M \times 1}}$ represents the beamforming vector used at BS $b$ for symbol ${x_{b,\left(l,i\right)}[t]}$. Also,
The received noise at the $i$-th user served by the $l$-th BS at the time step $t$ is ${n_{\left(l,i\right)}[t]}\sim \mathcal{CN}\left( {0,{\sigma _n}^2} \right)$. Finally, the received signal ${y_{\left(l,i\right)}[t]}$ will be used to compute the received signal-to-interference and noise ratio (SINR) ${\gamma _{\left( l,i\right)}}[t]$ of the $i$-th user served by $l$-th BS at the time step $t$.



\subsection{Problem Formulation}
The objective in this paper is, under some reasonable constraints, to gain the maximum value of the user equipment sum rate by jointly optimizing the beamforming vectors and transmit powers at the base stations based on the time-varying downlink model introduced before.
Here the formulation of joint power control, interference coordination and joint beamforming optimization problem is given by
\begin{equation}
\begin{aligned}
\mathop {\rm{maximize}}_{\substack{{P_{l,\left(l,i\right)}[t]},\forall l \\ {{\bf f}_{l,\left(l,i\right)}[t]},\forall l,i}}\qquad
&\sum_{l=1}^{L} \sum_{i=1}^{d_l}{\gamma _{\left( l,i\right)}}[t] \\
\rm{subject} \ \rm{to}\qquad & {P_{b,\left(l,i\right)}[t]}\geq 0, \quad &\forall b, l, i\\
&\sum_{\left( l,i\right)} {P_{b,\left(l,i\right)}[t]} \leq P_b, \quad &\forall b\\
&{{\bf f}_{b,\left(l,i\right)}}[t] \in {\bf F},\quad &\forall b, l,i\\
&{\gamma _{\left( l,i\right)}}[t] \ge {\gamma _{{\rm{cutoff}}}}, \quad &\forall l,i,
\end{aligned}
\end{equation}
where $P_b$ represents the maximum available transmit power at BS $b$, ${\bf F}$ represents the code book of beamforming vectors from which ${\bf f}_{b,\left(l,i\right)}[t]$ is chosen, and ${\gamma _{{\rm{cutoff}}}}$ denotes the cut-off value of SINR to ensure the efficient communication.

Concretely, each beamforming vector is selected from a BF code book ${\bf F}$ with cardinality $|F|: = {N_{CB}}$ based on the beamsteering and with the $n$-th element ${{\bf f}_n}$ following \cite{18}:
\begin{equation}
\begin{split}
\label{equation:empirical}
{{\bf f}_n}&:= a\left( \theta  \right)\\
&=\frac{1}{{\sqrt M }}{\left[ {1,{e^{jkd\cos \left( {{\theta _n}} \right)}},\cdots,{e^{jkd\left( {M - 1} \right)\cos \left( {{\theta _n}} \right)}}} \right]^T}
\end{split}
\end{equation}
where $d$ represents the antenna spacing and $k$ is the wave-number, and ${\theta _n}$ denotes the steering angle which is the division of the antenna angular spacing from 0 to $\pi $ radians by the amount of antennas $M$. Note that, the weights of each beamforming vector ${\bf f}_{b,\left(l,i\right)}[t]$ is produced by the constant-modulus phase shifters, namely ${[{{\bf f}_{b,\left(l,i\right)}[t]}]_m} = \frac{1}{{\sqrt M }}{e^{j{\theta _m}}}$.

\section{Preliminary Knowledge of Deep Reinforcement Learning}\label{Preliminary}
In this section, we briefly introduce the basic knowledge of DRL and several fundamental algorithms of our proposed algorithms.
\subsection{The Overview of DRL}
Reinforcement learning is one of typical methodologies in the field of machine learning, focusing on finding the optimal policy by learning the experience from the interaction with environment. In the meanwhile, deep learning is an advanced technique, which simulates the structure of human nervous system and imitates the way of processing information and improving the action to establish the neural network. By combining the perceptual ability of RL and the approximation ability of DL, deep reinforcement became one preferable method in artificial intelligence.

\begin{figure}[!t]
	\centering 
	\includegraphics[width=3in,height=2in]{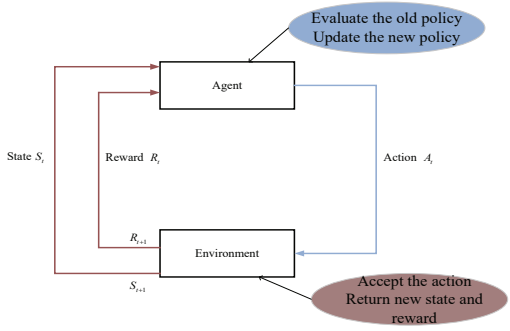}
	\caption{The interaction of reinforcement learning between the agent and environment.}
	\label{Fig.2}
\end{figure}

In order to build one RL model, there are several necessary elements as follows:
\begin{itemize}
	\item Agent: An agent is one independent entity that can observe and percept the environment and take actions accordingly in order to achieve its goal.
	\item Environment: The environment means the abstraction of the properties of machine learning problems, which interacts with the agents.
	\item State: A state ${S_t} \in S$ is what agent observes at time step \textit{t}, which can be used as the representation of environment.
	\item Action: An action ${A_t} \in A\left( {{S_t}} \right)$ is produced after agent percepts the state at time step \textit{t} according to the policy $\pi$ and can lead to the next state ${S_{t + 1}}$.
	\item Reward: A reward ${R_{t + 1}} \in R$ is the feedback from the environment after the agent takes one action ${A_t}$ in the state ${S_t}$ at time step \textit{t}.
	\item Policy: A policy $\pi \left( {\rm{\cdot}} \right)$ is one kind of guideline used by the agent to take actions, which maps the states to the actions.
	\item State-action value function: The state-action value function is the expectation of the rewards after taking one specific action by following the policy $\pi$, which is denoted by ${Q_\pi }\left( {s,a} \right)$ with
\begin{equation}
	{Q_\pi }\left( {s,a} \right) = E[{R_{t + 1}} + \alpha Q\left( {{S_{t + 1}},{A_{t + 1}}} \right)|{S_t} = s,{A_t} = a]
\end{equation}
	based on the Bellman function. Here $\alpha$ represents the discount factor.
\end{itemize}

With the interaction between agent and environment at discrete time steps, the agent gets the state at step \textit{t} and decides to take one action according to policy $\pi$. After this operation, one reward and the next state are returned to the agent by the environment. The process is shown in Fig. \ref{Fig.2}.

%

\begin{figure*}[!t]
		\centering 
		\includegraphics[width=7in,height=5in]{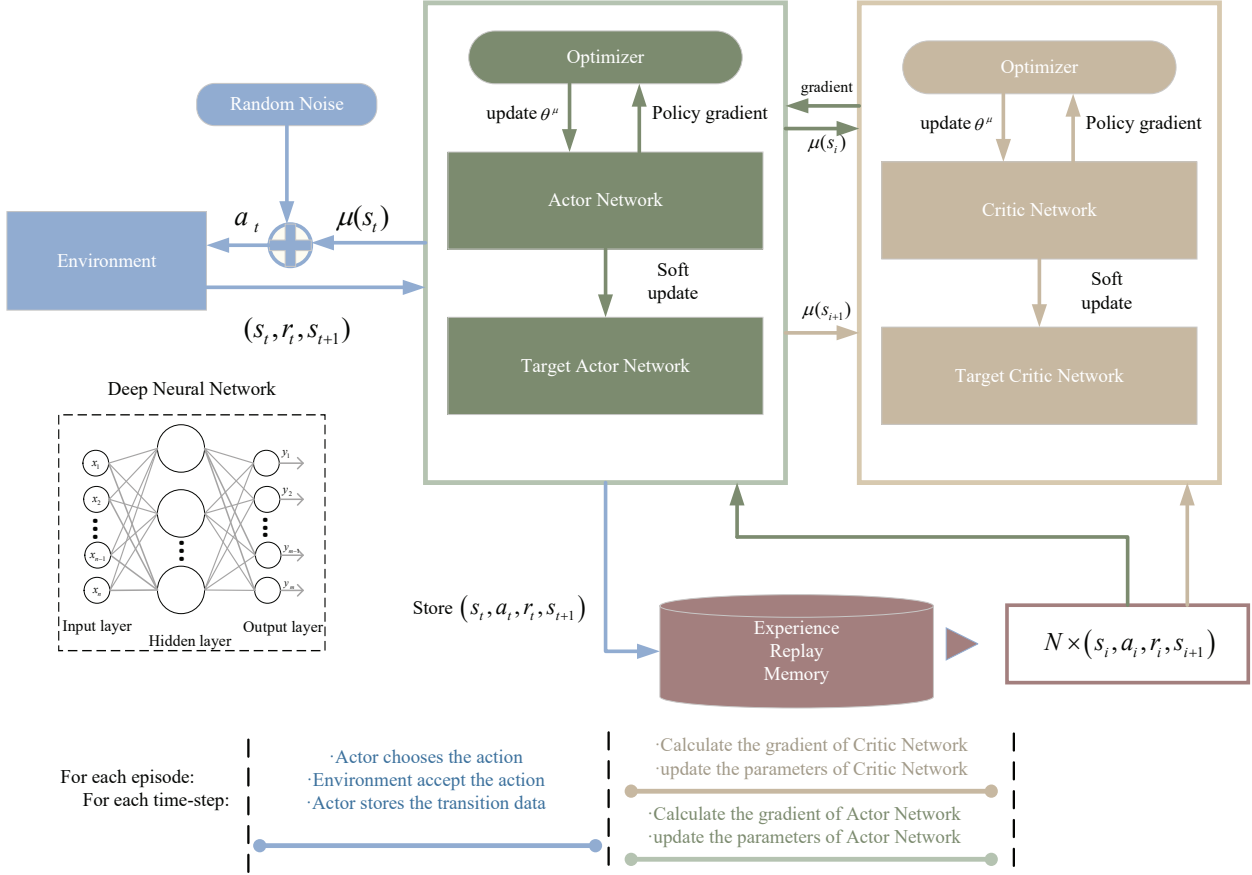}
		\caption{The architecture of DDPG.}
		\label{Fig.3}
	\end{figure*}

\subsection{The Fundamental Algorithms of DDPG}
DDPG is a deterministic algorithm based on Actor-Critic and Deep Q-learning. Actor and Critic in DDPG establish the approximate functions by using deep neural networks and back up a set of parameters as target network to calculate state-action value, and the policy from Actor produces a deterministic action.

\subsubsection{Deep Q-learning}DQN is a value-based and off-policy machine learning technique combining RL with DL theories. In order to reduce the complexity and the data correlation, it calculates the value of the state-action function $Q\left( \theta  \right)$ by deep neural network and optimizes by using experience replay.
\par It should be noted that agent can remember the previous experience of state transition in the experience buffer. For each state transition in every complete state sequence, agent takes one action ${a_t}$ with $\varepsilon $-greedy policy according to current state ${s_t}$, gets the corresponding reward ${r_{t + 1}}$ and the next state ${s_{t + 1}}$, and deposits this transition $\left( {{s_t},{a_t},{r_t},{s_{t+1}}} \right)$ in the memory. When the memory capacity is large enough, mini-batch size state transitions are extracted at random in order to calculate the target state-action values ${Q_{{\rm{target}}}}\left( {{s_t},{a_t}} \right)$ of the current state-action pairs by using next pairs in transitions:
\begin{equation}
{Q_{{\rm{target}}}}\left( {{s_t},{a_t}} \right) = {r_t} + \alpha {\rm{max }}Q\left( {{s_{t+1}},{a_{t+1}};{\theta ^ - }} \right)
\end{equation}
where $\theta$ represents the parameter of the DNN and $\alpha$ denotes the discount factor.

After that, the target values are used in the loss function to perform the mini-batch gradient descent. At last, the parameters in deep neural network are updated.

\subsubsection{Actor-Critic}
Actor-Critic is a policy-based algorithm for temporal difference tasks, including a policy function called Actor to interact with the environment and to produce the actions, and a state-action value function called Critic to evaluate the performance of Actor.

The state-action value function of Critic is an approximation based on the policy ${\pi _\theta }$:
\begin{equation}
{Q_\omega }\left( {s,a} \right) \approx {Q_{{\pi _\theta }}}\left( {s,a} \right)
\end{equation}
Where $\omega $ represents the parameter of the state-action value function.

Simple Actor-Critic algorithm does not require complete state sequence. However, since Critic is an approximate function that may cause bias, the policy gradient is not accurate if not lucky enough.

\section{DDPG Based on Joint Beamforming, Power and Interference Coordination}\label{DDPG}
In this section, we describe the details of our first proposed scheme, i.e., DDPG based on joint beamforming, power control and interference coordination.

\subsection{The Overview of DDPG Algorithm}
There are four fundamental neural networks in DDPG shown in Fig.\ref{Fig.3} such as actor network, target actor network, critic network and target critic network as follows.

\begin{itemize}
	\item Actor network: It produces a specific action ${a_t}$ according to the current state ${s_t}$.
	\item Target Actor network: It produces the next action ${a_t}'$ to calculate the predicted value according to the next state from the environment.
	\item Critic network: It calculates the state-action value corresponding to the state ${s_t}$ and the action ${a_t}$.
	\item Target Critic network: It calculate the state-action value $Q'\left( {{s_{t+1}},{a_{t+1}}} \right)$ corresponding to the next state ${s_{t + 1}}$ and action ${a_{t + 1}}$ in order to calculate the target value $y = Q\left( {{s_t},{a_t}} \right)$.
\end{itemize}

\par DDPG remedies both Deep Q-learning and Actor-Critic:
\subsubsection{Continuous Output} 
\begin{figure}[!t]
	\centering 
	\includegraphics[width=3in,height=2in]{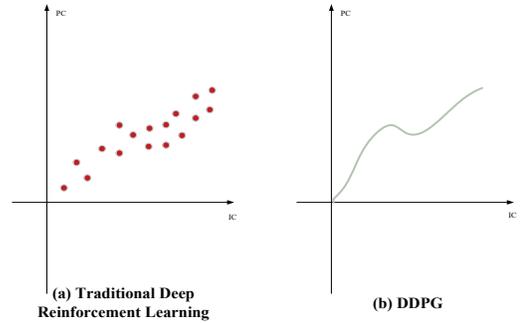}
	\caption{The comparation of discrete and continuous outputs of joint PC and IC. (a)discrete output by using traditional DRL; (b)continuous output by DDPG }
	\label{Fig.continuous}
\end{figure}

On the one hand, traditional deep reinforcement learning method such as DQN can only solve the low dimensional tasks with discrete action outputs, which leads that  some useful information may be abandoned since DQN cannot the direct output of continuous actions. However, the application of Deterministic Policy Gradient in DDPG make outputting continuous actions possible shown in Fig.\ref{Fig.continuous} . This is because DDPG regards the policy as an approximation function achieved by the deep neural network, which is continuous in the mapping domain and can just get one deterministic output. In the meanwhile, the exploration in DDPG is through the addition of random noise and the output policy.

\subsubsection{Experience Replay} On the other hand, it is not easy for Actor-Critic to converge since the actions from Actor depend on the state-action values from Critic. In order to solve this problem, DDPG absorbs the advantages of DQN to use off-policy method, which can minimize the correlation among samples by sampling in the replay buffer.

\par In summary, DDPG introduced before can output continuous actions and produce the deterministic one, which is more in line with actual situations of base stations. Also, It not only has the same complexity at the order of magnitude as DQN, but also improves the performance.

\subsection{DDPG based joint PC, IC and BF}
According to the practical situations and preliminary knowledge of DDPG introduced before, we choose some necessary information as the states and the index of BF, continuous PC and IC actions.

\subsubsection{State} As for the setup of states, there are some details as follows:
\par $s_t^0$: the \textit{x} position of the UE served by BS \textit{l} at the time \textit{t}.
\par $s_t^1$: the \textit{y} position of the UE served by BS \textit{l} at the time \textit{t}.
\par $s_t^2$: the \textit{x} position of the UE served by BS \textit{b} at the time \textit{t}.
\par $s_t^3$: the \textit{y} position of the UE served by BS \textit{b} at the time \textit{t}.
\par $s_t^4$: the transmit power of BS \textit{l}.
\par $s_t^5$: the transmit power of BS \textit{b}.
\par $s_t^6$: the index of beamforming vector code book for BS \textit{l}.
\par $s_t^7$: the index of beamforming vector code book for BS \textit{b}.

\subsubsection{Action}
\par The power control ${P_{l}}[t]$ for the serving BS \textit{l} at time step \textit{t} is as follows:
\begin{equation}
{P_{l}}[t] = \min \left( {P_{BS}^{\max },PC_l[t]} \right)
\end{equation}
where ${P_{BS}^{\max }}$ means the maximum value of one base station's transmit power and $PC_l[t]$ represents the transmit power of BS \textit{l} after power control.
\par The interference coordination on the interfering BS \textit{b} at time step \textit{t} can be written as:
\begin{equation}
{P_{b}}[t] = \min \left( {P_{BS}^{\max },IC_b[t]} \right)
\end{equation}
where $IC_b[t]$ represents the transmit power of BS \textit{b} after interference coordination.
\par As for the data, the special new action about stepping up or down the beamforming for the BSs \textit{b} and \textit{l} independently is to use circular increments or decrements $(\textit{n}\pm 1)$. And DDPG achieves the selection of \textit{n} by rounding down the continuous output:

\begin{equation}
n \mapsto {\textbf{f}_n}[t]:n: = \left( {n \pm 1} \right)\bmod M
\end{equation}

Therefore, the settings of actions are as follows:
\par $a_t^0$:the value of ${P_{l}}$ after the continuous change
\par $a_t^1$:the value of ${P_{b}}$ after the continuous change
\par $a_t^2$:the index of beamforming vector code book for BS \textit{l}.
\par $a_t^3$:the index of beamforming vector code book for BS \textit{b}.

\subsubsection{Reward}The setting of reward is related to the goal and can be expressed as follows:
\begin{equation}
{r_{s,s',a}}[t] =  \sum_{i=1}^{d_l}{\gamma _{\left( l,i\right)}}[t] +   \sum_{b \neq l}\sum_{i=1}^{d_b}{\gamma _{\left( b,i\right)}}[t]
\end{equation}
where ${r_{s,s',a}}[t]$ is the reward from the state \textit{s} to the next state \textit{s}' at time step \textit{t} after taking the action \textit{a}.

\renewcommand{\algorithmicrequire}{\textbf{Input:}}
\renewcommand{\algorithmicensure}{\textbf{Output:}}
\begin{algorithm}
	\caption{DDPG}
	\begin{algorithmic}[1]
		\REQUIRE $\alpha $, $\tau $, ${\theta ^Q}$, ${\theta ^\mu }$
		\ENSURE optimized ${\theta ^Q}$, ${\theta ^\mu }$
		\STATE Randomly initialize critic network $Q\left( {s,a|{\theta ^Q}} \right)$ and actor network $\mu \left( {s|{\theta ^\mu }} \right)$ with weights ${\theta ^Q}$ and ${\theta ^\mu }$
		\STATE Initialize target network $Q'$ and $\mu '$ with weights ${\theta ^{Q'}} \leftarrow {\theta ^Q}$, ${\theta ^{\mu '}} \leftarrow {\theta ^\mu }$
		\STATE Initialize replay experience buffer \textit{R}
		\FOR {episode from 1 to \textit{Limit}}
		\STATE Initialize a random process(noise) \textit{N} for action exploration
		\STATE Receive initial observation state ${s_1}$
		\FOR {t=1 to \textit{T}}
		\STATE Select action ${a_t} = \mu \left( {{s_t}|{\theta ^\mu }} \right) + N{}_t$ according to the current policy and exploration noise
		\STATE Execute action ${a_t}$, observe reward ${r_{t + {\rm{1}}}}$ and new state ${s_{t + {\rm{1}}}}$
		\STATE Store transition $\left( {{s_t},{a_t},{r_{t + 1}},{s_{t + 1}}} \right)$ in \textit{R}
		\STATE Sample a random minibatch of \textit{M} transitions $\left( {{s_i},{a_i},{r_{i + 1}},{s_{i + 1}}} \right)$ from \textit{R}
		\STATE set ${y_i} = {r_{i + 1}} + \alpha Q'\left( {{s_{i + 1}},\mu '({s_{i + 1}}|{\theta ^{\mu '}})|{\theta ^{Q'}}} \right)$
		\STATE Update critic by minimizing the loss:
		\STATE $L = \frac{{\rm{1}}}{M}\sum\limits_i {{{({y_i} - Q({s_i},{a_i}|{\theta ^{Q'}}))}^2}} $
		\STATE Update the actor policy using the sampled policy gradient:
		\STATE ${\nabla _{{\theta ^\mu }}}J \approx \frac{{\rm{1}}}{M}{\nabla _a}Q(s,a|{\theta ^Q}){|_{s = s,a = \mu (s)}}{\nabla _{{\theta ^\mu }}}\mu (s|{\theta ^\mu }){|_{{s_i}}}$
		\STATE Update the target networks:
		\STATE ${\theta ^{Q'}} \leftarrow \tau {\theta ^Q} + (1 - \tau ){\theta ^{Q'}}$
		\STATE ${\theta ^{\mu '}} \leftarrow \tau {\theta ^\mu } + (1 - \tau ){\theta ^{\mu '}}$
		\ENDFOR
		\ENDFOR
	\end{algorithmic}
\end{algorithm}

\par And the cut-off value of SINR are set as 4dB, which is the maximum value that can ensure the success in all situations after many experiments since it means the episode should be aborted if the SINR is less than it in order to increase the performance and ensure the completeness of data transmission. At last, the details of process are shown in Algorithm 1.

\subsection{Construction of DNN}
\begin{figure}[!t]
	\centering 
	\includegraphics[width=2.5in,height=1.67in]{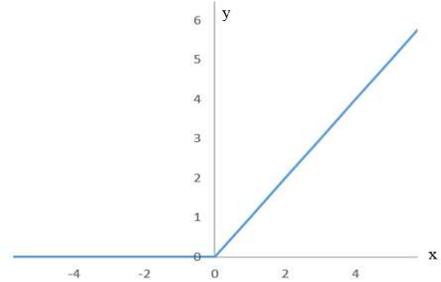}
	\caption{The diagram of \textit{ReLU} function.}
	\label{Fig.5}
\end{figure}

It should be noted that we abandoned \textit{ReLU} but selected \textit{tanh} as the activation function of DDPG. Although \textit{ReLU} is the most commonly used activation function, there is one main disadvantage of it causing the relatively terrible performance in our model, which is Dead \textit{ReLU} problem. As the diagram shown in Fig. \ref{Fig.5}, the output of \textit{ReLU} is not zero-centered and some neurons may never be activated, resulting in the corresponding parameters to never be updated.

\par The expression of \textit{ReLU} function can be written as:
\begin{equation}
{\rm{Relu}} = {\rm{max(0,x)}}
\end{equation}
\par
\begin{figure}[!t]
	\centering 
	\includegraphics[width=2.5in,height=1.67in]{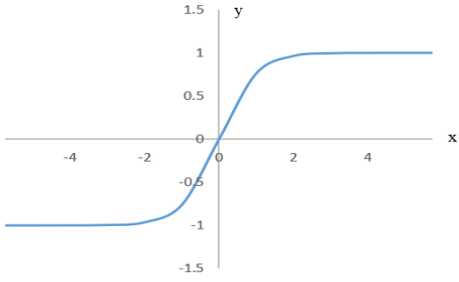}
	\caption{The diagram of \textit{tanh} function.}
	\label{Fig.6}
\end{figure}

 This defect performs seriously in our model in experiments of DDPG and h-DDPG. In order to improve the performance, we choose \textit{tanh} function\cite{36}. As the diagram shown in Fig. \ref{Fig.6}, it can be represented as:

\begin{equation}
\tanh (x) = \frac{{{e^x} - {e^{ - x}}}}{{{e^x} + {e^{ - x}}}}
\end{equation}

\par It solves the non-activation of neurons to some degrees but due to the introduction of exponential functions, the gradient vanishing problem exists. Therefore, we increase the value of minimum batch size to 128, which makes the gradient accurate and valid.

\section{Hierarchical DDPG Based Joint Beamforming, Power Control and Interference Coordination} \label{Hiera_DDPG}

In this section, we introduce the hierarchical theory and our second proposed algorithm, Hierarchical DDPG, based joint beamforming, power control and interference coordination.
\subsection{Hierarchical DDPG}
Since deep reinforcement learning is an end-to-end algorithm, it is important for agent to receive enough positive rewards in order to ensure the effectiveness of learning, which means the long-timescale sparse reward problem became a bottleneck for DRL.
\par Hierarchical theory aims at decomposing a complex DRL problem into several sub-problems and solve them separately. In this paper, we focus on the idea of meta-policy, which divides the agent into two layers, meta controller and controller:

\begin{figure*}[!t]
	\centering 
	\includegraphics[width=7in,height=5in]{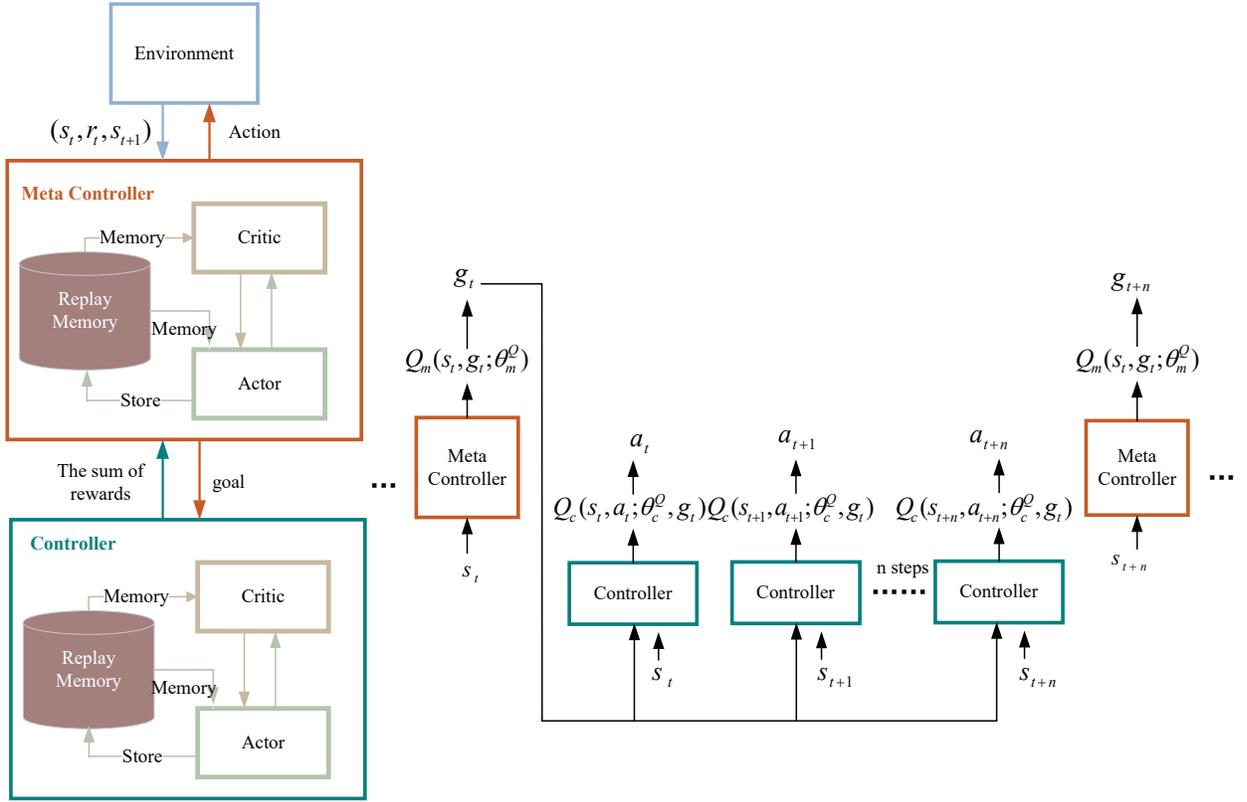}
	\caption{The architecture of hierarchical DDPG.}
	\label{Fig.4}
\end{figure*}
	\subsubsection{Meta controller}The responsibility of one meta controller is to set a goal so that the controller can achieve it and the goals are equivalent to the action at time \textit{t}. In the meanwhile, the state space of meta controller is still the original one. Different from the agent in an ordinary DRL, the time scale of meta controller is coarser, which means one time step of it can be counted as several normal time steps. It makes rewards not sparse in the view of meta controller.
	
\subsubsection{Controller}Controller takes actions at each time step \textit{t} in every episode to achieve the goal set by meta controller, which is similar to the ordinary agent.
\par Meta controller and controller respectively have one policy system and the corresponding state-action value system. Throughout the sub processes, the agent observes the environment at the beginning and then meta controller set one goal according to its policy function. After it, during one time step of meta controller, namely $c$ normal time steps, controller takes action like an ordinary DRL agent to achieve the goal. And the whole process is composed of repeated sub processes.
\par In this paper, we applies meta-policy theory based on DDPG to pursue the better performance of the model. Meta controller and controller respectively have one DDPG system to learn from the observation and the architecture is shown in Fig. \ref{Fig.4}.

\subsection{Hierarchical DDPG Based Joint PC, IC and BF}
Applying the hierarchical theory on the DDPG, to some degree, we solve the sparse reward problem due to the abortion of the unsuccessful episodes in this model.
\subsubsection{State and Action}
The setup of state and action of controller in hierarchical DDPG is the same as DDPG since the information needed from the environment does not change due to the introduction of hierarchical theory and the form of joint actions are still continuous according to the application of DDPG.
\subsubsection{Goal }
As the introduction before, goal is similar to the concept of action in ordinary reinforcement learning. In order to keep the consistency with controller, here are the settings of goals of meta controller as follows.
\par $g_t^0$:the value of ${P_{l}}$ after the continuous change
\par $g_t^1$:the value of ${P_{b}}$ after the continuous change
\par $g_t^2$:the index of beamforming vector code book for BS \textit{l}.
\par $g_t^3$:the index of beamforming vector code book for BS \textit{b}.

\subsubsection{Reward}
\par As for the reward ${r_{s,s',a}}[t]$ of the controller at time step \textit{t}, it should be related to not only the final objective but also the goals set by the meta controller. Therefore, it can be expressed as:
\begin{equation}
{r_{s,s',a}}[t] = \sum_{i=1}^{d_l}{\gamma _{\left( l,i\right)}}[t] +   \sum_{b \neq l}\sum_{i=1}^{d_b}{\gamma _{\left( b,i\right)}}[t] - (goal - action)
\end{equation}
\par Note that, the subtraction of goal and action can be regarded as the sum of the absolute value of the corresponding elements' substraction.
\par As for the reward of meta controller ${R_{s -c + 1,s',a}}[t]$ at time step \textit{t}, it can be represented as:
\begin{equation}
{R_{s -c + 1,s',a}}[t] = \sum\limits_{i = s - c + 1}^{t + 1} {r[i]}
\end{equation}
where \textit{s-c+1} means the starting sate of meta controller and \textit{s}' is the current state after fixed $c$ steps.

\renewcommand{\algorithmicrequire}{\textbf{Input:}}
\renewcommand{\algorithmicensure}{\textbf{Output:}}
\begin{algorithm}
	\caption{h-DDPG}
	\begin{algorithmic}[1]
		\REQUIRE $\alpha $, $\tau $, $\theta _c^Q$, $\theta _c^\mu $, $\theta _m^Q$, $\theta _m^\mu $, \textit{c}
		\ENSURE optimized $\theta _c^Q$, $\theta _c^\mu $, $\theta _m^Q$, $\theta _m^\mu $
		\STATE Randomly initialize critic network ${Q_c}\left( {s,a|{\theta _c}^Q} \right)$ and actor network ${\mu _c}\left( {s|{\theta _c}^\mu } \right)$ with weights $\theta _c^Q$ and $\theta _c^\mu $ for the controller
		\STATE Randomly initialize critic network ${Q_m}\left( {s,a|{\theta _m}^Q} \right)$ and actor network ${\mu _m}\left( {s|{\theta _m}^\mu } \right)$ with weights $\theta _m^Q$ and $\theta _m^\mu $ for the meta controller
		\STATE Initialize target network ${Q_c}'$ and ${\mu _c}'$ with weights $\theta _c^{Q'} \leftarrow \theta _c^Q$, $\theta _c^{\mu '} \leftarrow \theta _c^\mu $
		\STATE Initialize target network ${Q_m}'$ and ${\mu _m}'$ with weights $\theta _m^{Q'} \leftarrow \theta _m^Q$, $\theta _m^{\mu '} \leftarrow \theta _m^\mu $
		\STATE Initialize replay experience buffer \textit{R}, \textit{B}
		\FOR {episode from 1 to \textit{Limit}}
		\STATE Initialize a random process(noise) \textit{N} for action exploration
		\STATE Receive initial observation state ${s_1}$
		\STATE Select goal ${g_t} = {\mu _m}\left( {{s_t}|{\theta _m}^\mu } \right) + N{}_t$ from meta controller according to the current policy and exploration noise
		\FOR {t=1 to \textit{T}}
		\STATE Select action ${a_t} = {\mu _c}\left( {{s_t}|{\theta _c}^\mu } \right) + N{}_t$ according to the current policy and exploration noise
		\STATE Execute action ${a_t}$, observe reward ${r_{t + {\rm{1}}}}$ and new state ${s_{t + {\rm{1}}}}$
		\STATE Store transition $\left( {{s_t},{a_t},{r_{t + 1}},{s_{t + 1}}} \right)$ in \textit{R}
		\STATE Sample a random minibatch of \textit{M} transitions $\left( {{s_i},{a_i},{r_{i + 1}},{s_{i + 1}}} \right)$ from \textit{R}
		\STATE set ${y_i} = {r_{i + 1}} + \alpha {Q_c}'\left( {{s_{i + 1}},{\mu _c}'({s_{i + 1}}|{\theta _c}^{\mu '})|{\theta _c}^{Q'}} \right)$
		\STATE Update critic by minimizing the loss:
		\STATE $L = \frac{{\rm{1}}}{M}\sum\limits_i {{{({y_i} - {Q_c}({s_i},{a_i}|{\theta _c}^{Q'}))}^2}} $
		\STATE Update the actor policy using the sampled policy gradient:
		\STATE ${\nabla _{{\theta _c}^\mu }}J \approx \frac{{\rm{1}}}{M}{\nabla _a}Q_c^{}(s,a|{\theta _c}^{{Q}}){|_{s = s,a = {\mu _c}(s)}}{\nabla _{{\theta _c}^\mu }}{\mu _c}(s|{\theta _c}^\mu ){|_{{s_i}}}$
		\STATE Update the target networks:
		\STATE ${\theta _c}^{Q'} \leftarrow \tau {\theta _c}^Q + (1 - \tau ){\theta _c}^{Q'}$
		\STATE \[{\theta _c}^{\mu '} \leftarrow \tau {\theta _c}^\mu  + (1 - \tau ){\theta _c}^{\mu '}\]
		\IF{t mod c is 0}
		\STATE Calculate the reward ${R_{t + 1}} = \sum\limits_{i = t - c + 1}^{t + 1} {{r_i}} $ for the meta controller
		\STATE Store transition $\left( {{s_{t{\rm{ - }}c + 1}},{g_{t - c + 1}},{R_{t + 1}},{s_{t + 1}}} \right)$ in \textit{B}
		\STATE Repeat the minibatch replay and update the parameters $\theta _m^Q$, $\theta _m^\mu $ and actor policy
		\ENDIF
		\ENDFOR
		\ENDFOR
	\end{algorithmic}
\end{algorithm}
\par As for the activation function, since controller and meta controller use respectively independent DDPG system, the settings are the same as DDPG introduced before. And the minimum SINR are set as 4dB, which is the maximum values that can ensure the success in all situations after many experiments with the same reasons as DDPG.
\par And the settings of power control and interference coordination follows DDPG. Also, hierarchical DDPG achieves the selection of \textit{n} by rounding down the continuous output. The detailed process of the hierarchical DDPG based algorithm is shown in Algorithm 2.

\section{Simulation and Results} \label{Simulations}
In this section, we present the comparative algorithm and simulation details, performance measures, and some numerical results.
\subsection{Comparison Algorithms and Simulation Details}
In order to show the performance improvement of our proposed algorithm, it is necessary to set proper comparison algorithms and the parameters of simulation.
There are three algorithms with discrete action output as follows.
\subsubsection{Comparison Algorithms }
\begin{itemize}
	\item Fixed Power Allocation: Fixed Power Allocation (FPA) is a standard algorithm applied in the industry at present, which aims at the simplest situation in our model. The transmit powers of the serving base station is dedicated and the interference coordination is not included, which causes that the link adaption is achieved depending on feedback from the periodic or aperiodic measurement of UE. FPA just changes the code schemes of transmission and the modulation.
	The transmit power of the \textit{a}-th base station at time \textit{t} is represented as ${P_{a}}[t]$:
	\begin{equation}
	{P_{a}}[t]:= P_{BS}^{\max } - 10\log {N_{PRB}} + 10\log {N_{PRB,a}}[t] \ (\rm dBm)
	\end{equation}
	where ${N_{PRB}}$  is the total number of physical resource blocks in BS and ${N_{PRB,a}}[t]$ denotes the number of available PRBs at the \textit{b}-th base station at the time step \textit{t}. And $P_{BS}^{\max }$ represents the maximum value of one base station's transmit power. The value of this formula is constant since it is divided equally among all the physical resource blocks in BS. FPA can improve the effective SINR received by UE and reduce the voice packet error rate. The main issue is the requirement about the feedback from UE.
	
	\item Q-learning: Q-learning is a traditional off-policy reinforcement learning technique and there is no neural network in it, which mainly focuses on voice bearers.
	The update of the state-action function $Q\left( {{s_t},{a_t}} \right)$ is as follows:
	\begin{equation}
	\begin{aligned}
	Q\left( {{s_t},{a_t}} \right)= &Q\left( {{s_t},{a_t}} \right)\\
	& + \lambda \left( {{r_{t + 1}} + \alpha \mathop {\max }\limits_{a'} Q\left( {{s_{t + 1}},a'} \right) - Q\left( {{s_t},{a_t}} \right)} \right)
	\end{aligned}	
	\end{equation}
	
	where $\lambda  > 0$ is the learning rate and $\alpha $ is the discount factor.
According to this method, the state-action value of the current state ${s_t}$ and action ${a_t}$ from the $\varepsilon $-policy can be updated towards the direction of getting maximum state-action value in the next state ${s_{t + {\rm{1}}}}$. Therefore, Q-learning can perform without the frequent feedback from user equipment.
	
	\item DQN: As the introduction before, Deep Q-learning is a deep reinforcement learning method improved by Q-learning with neural network. In order to output the discrete actions, we set the range of increase and decrease each time for PC and IC is $ \pm {\rm{1}}$dB or $ \pm {\rm{3}}$dB and the setup of states is the same as our proposed algorithm. Also, the complexity of it reduces comparing with FPA and Q-learning since DQN does not need to set and record a table.
\end{itemize}

\subsubsection{Simulation Details}
For the deep reinforcement learning, the selections of parameters are under the careful consideration as shown in Table. \ref{table2}.

\begin{table}[!t]
	\renewcommand{\arraystretch}{1.3}
	\caption{The parameters of DDPG and h-DDPG.}
	\label{table2}
	\centering
	\begin{tabular}{|l|l|}
		\hline
		Parameter            & Value \\ \hline
		Discount factor $\alpha $     & 0.9   \\ \hline
		Number of the states $|{S}|$ & 8     \\ \hline
	Number of actions $(|{A}|)$ &(4 \\ \hline
	Soft update parameter $\tau $ & 0.1 \\ \hline
	Learning rate & 0.0001 \\ \hline
	DDPG width $H$ & 28 \\ \hline
	DDPG depth &4 \\ \hline
	Action limitations $({L_{PC}},{L_{IC}},{L_{BF}})$ & (40, 40, $k \times M-1$) \\ \hline
	Number of hierarchical steps ${c_{data}}$ & 3 \\ \hline
	DDPG Batch size ${N_{mb}}$ &128 \\ \hline
	h-DDPG Batch size $(N_{mb}^{meta},N_{mb}^{controller})$ &(64,64)\\ \hline
\end{tabular}
\end{table}
\par On the other hand, for the setups of the joint power control, interference coordination and beamforming environment shown in Table. \ref{table3}, the cellular network is adopted as the introduction before.

\begin{table}[!t]
\renewcommand{\arraystretch}{1.3}
\caption{The Parameters of the environment model.}
\label{table3}
\centering
\begin{tabular}{|l|l|}
	\hline
	Parameter           & Value \\ \hline
	Base station maximum transmit power $P_{BS}^{\max }$   & 40W   \\ \hline
	Downlink frequency band & (2100 MHz, 28GHz)     \\ \hline
	Cellular geometry &circular \\ \hline
	Cell radius \textit{r} & (350,150) m \\ \hline
	Propagation model data & \cite{37} \\ \hline
	User equipment (UE) antenna gain & 0dBi \\ \hline
	Antenna gain $G_{TX}^{data}$ &3 dBi \\ \hline
	Inter-site distance \textit{R} & (525,225) m\\ \hline
	Number of multipaths ${N_p}$ &(15,4)\\ \hline
	Probability of LOS $p_{LOS}$ &  0.8 \\ \hline
	UE average movement speed \textit{v} &(5,2) km/h\\ \hline
	Number of transmit antennas ${M^{data}}$ &$\{ 4,8,16,32,64\} $\\ \hline
	Radio frame duration ${T^{data}}$ &10 ms\\ \hline
\end{tabular}
\end{table}
\par At last, in order to plot the effective SINR, we set the target SINRs as:
\begin{equation}
\gamma _{{\rm{target}}}: = \gamma _{\rm{0}}+ {\rm{10}}\log M
\end{equation}
where $\gamma _{{\rm{target}}}$ denotes the target SINR, $M$ is the amount of the anennas and $\gamma _{\rm{0}}$ is  a constant threshold which is taken as 5dB in the simulation.

\subsection{Performance Measures}
There are several performance measures introduced to evaluate the algorithms as follows.
\subsubsection{Convergence}The convergence $\xi $ of the model in the learning process is defined as the convergence point of the loss values of models. Due to the abortion rule of episode, the reward or the state-action value with respect to the episodes may abruptly reduce to a very small value to influence our judgement of convergence point.
\subsubsection{Coverage}We defined the complement cumulative distribution function (CCDF) of $\gamma _{\rm{eff}}^l$, namely the probability of ${\gamma ^l}$ exceeding $\gamma _{\rm{cutoff}}^l$, which follows \cite{38}. Here $\gamma _{\rm{eff}}^l$ means the sum of effective SINR of all users served by the \textit{l}-th BS. The data is from the result of changing random seeds and running the experiment many times.
\subsubsection{Sum-rate Capacity} The function of average sum-rate capacity \textit{C} is as follows, which is the average of all received SINR from all base stations.
\begin{equation}
C = \frac{{\rm{1}}}{T}\sum\limits_{t = 1}^T {\sum\limits_{j \in \{ l,b\} } {{{\log }_2}(1 + \gamma _{\rm{eff}}^j[t])} }
\end{equation}
where \textit{l} represents the \textit{l}-th serving BS while \textit{b} is the \textit{b}-th interference BS. $\gamma _{\rm{eff}}^j[t]$ denotes the the sum of effective SINR of all users served by the \textit{l}-th BS at the time step \textit{t} from the \textit{j}-th BS and cannot be null value or exceed the maximum value. Here the maximum value follows \cite{18}:

\begin{equation}
{\gamma _{\max }} = {\gamma _0} + 10{\log _2}M
\end{equation}
where ${\gamma _0}$ is the constant threshold and \textit{M} denotes the number of antennas.
\par The settings of maximum value is because the ability of the receiver in UE to receive the signal is limited. It may produce the intermodulation distortion products when the receiver is overwhelmed with input signal. Even if the input signal is limited, we can improve the SINR by changing the interference. However, with the premise that the SINR of all UEs served by different BSs should be increased concurrently, interference coordination dose not work well but the increasing antenna gain with the growth of the number of antennas can play a role to some degree.

\subsection{Results}
\begin{figure}[!t]
\centering 
\includegraphics[width=3in,height=2in]{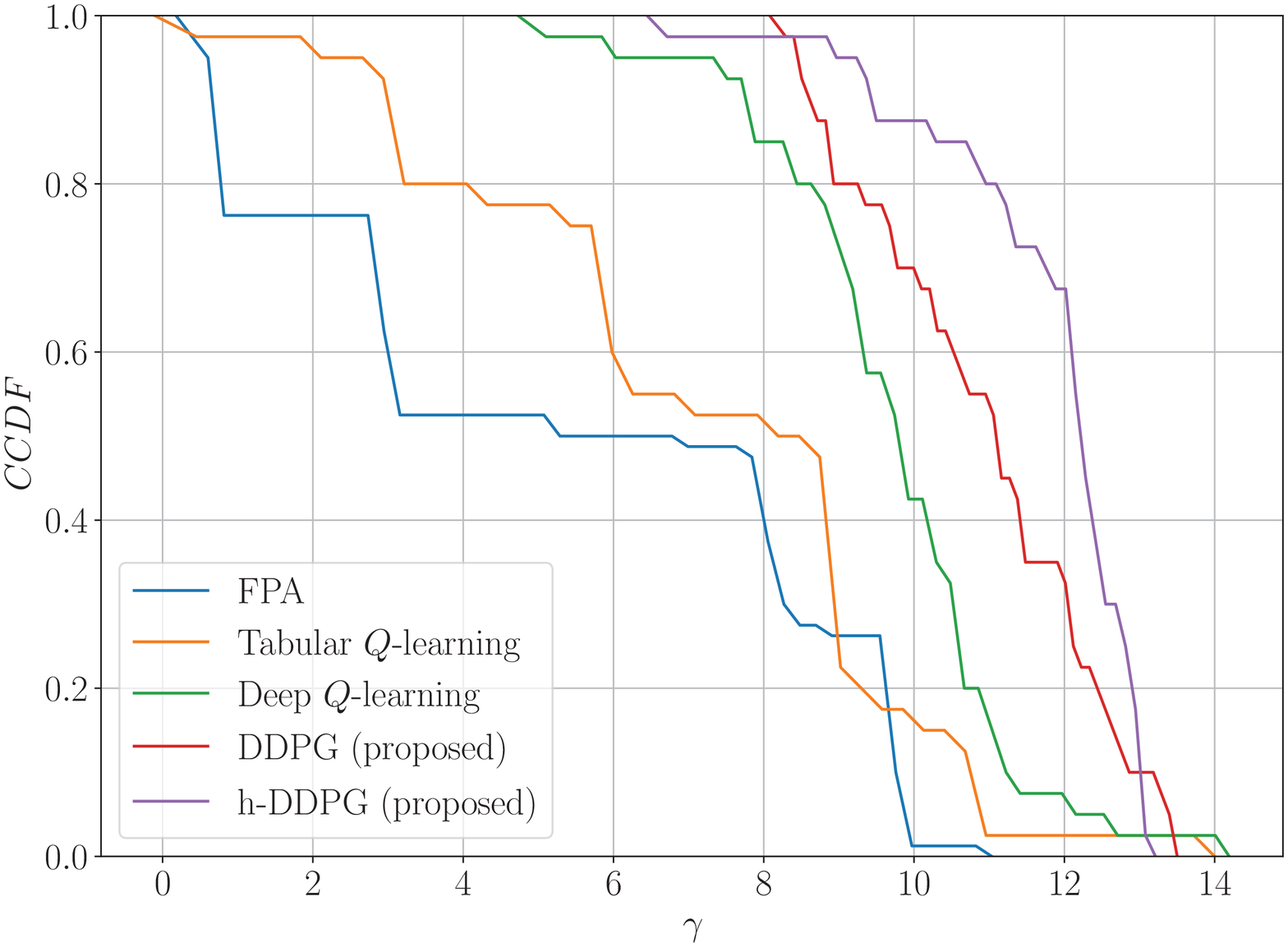}
\caption{The CCDF diagram of effective SINRs without beamforming.}
\label{Fig.7}
\end{figure}

\begin{figure*}[!t]
\centering
\subfloat[\textit{M}=4]{\includegraphics[width=2in,height=1.3in]{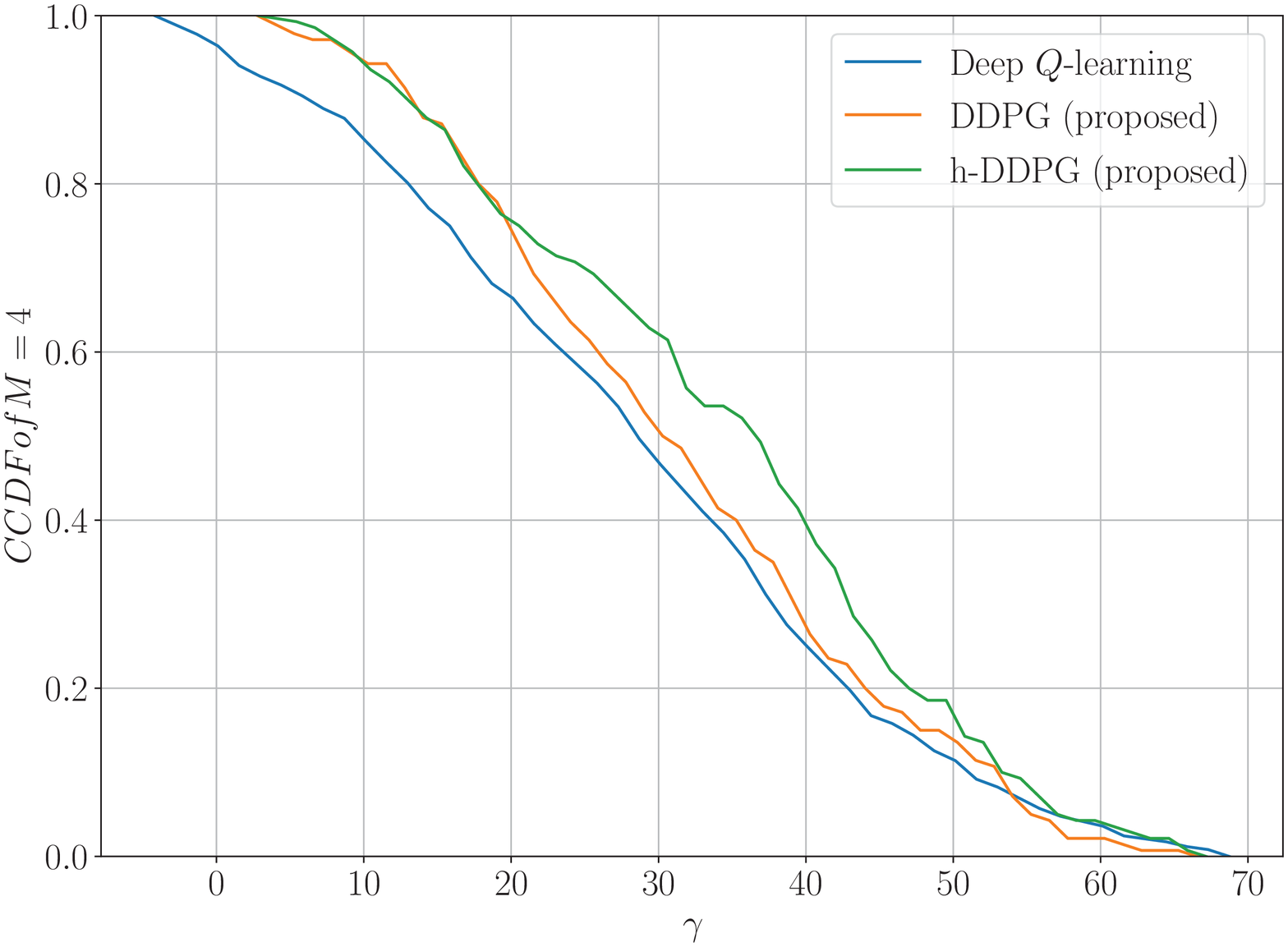}
}
\hfil
\subfloat[\textit{M}=8]{\includegraphics[width=2in,height=1.3in]{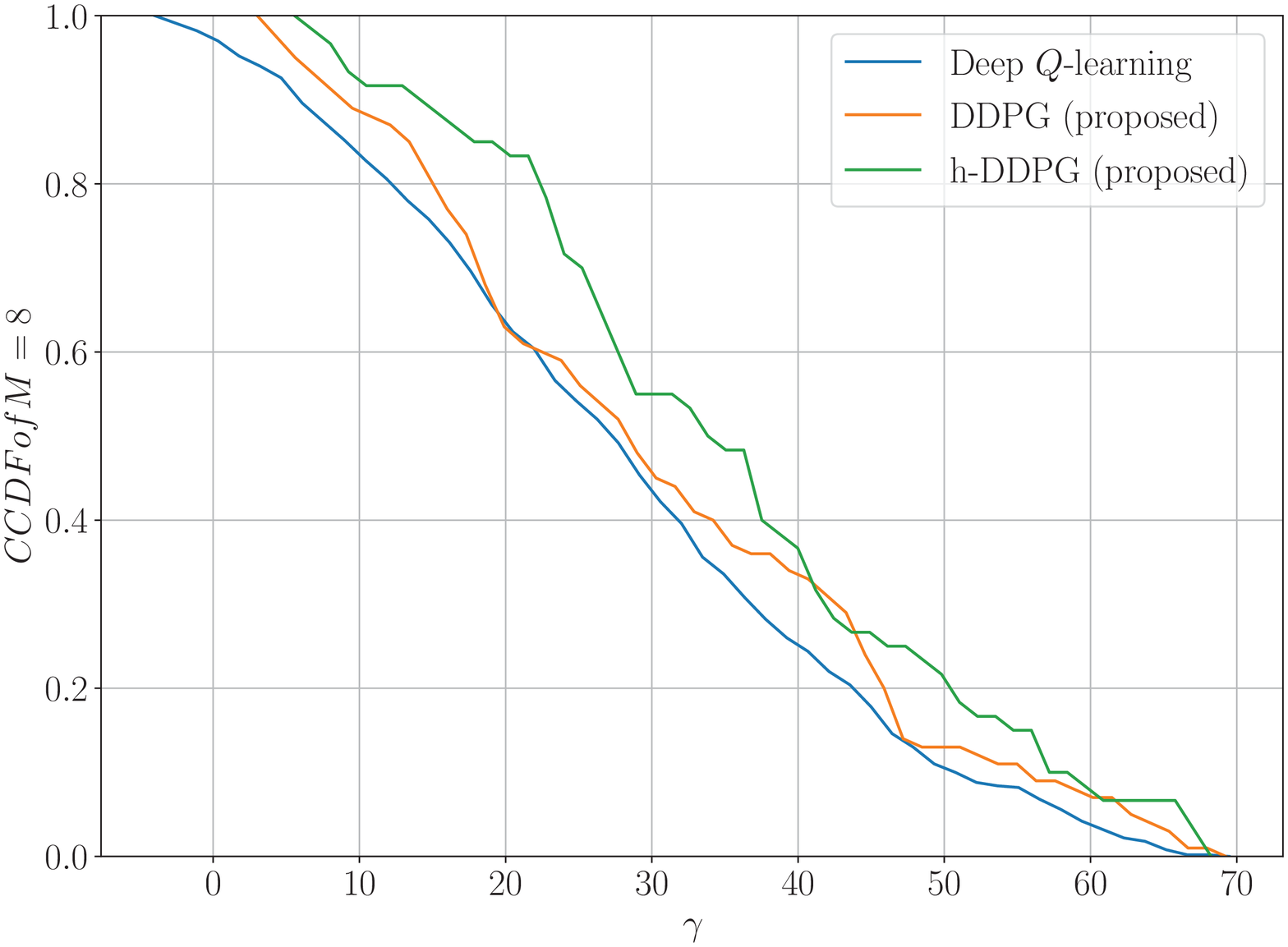}
}
\hfil
\subfloat[\textit{M}=16]{\includegraphics[width=2in,height=1.3in]{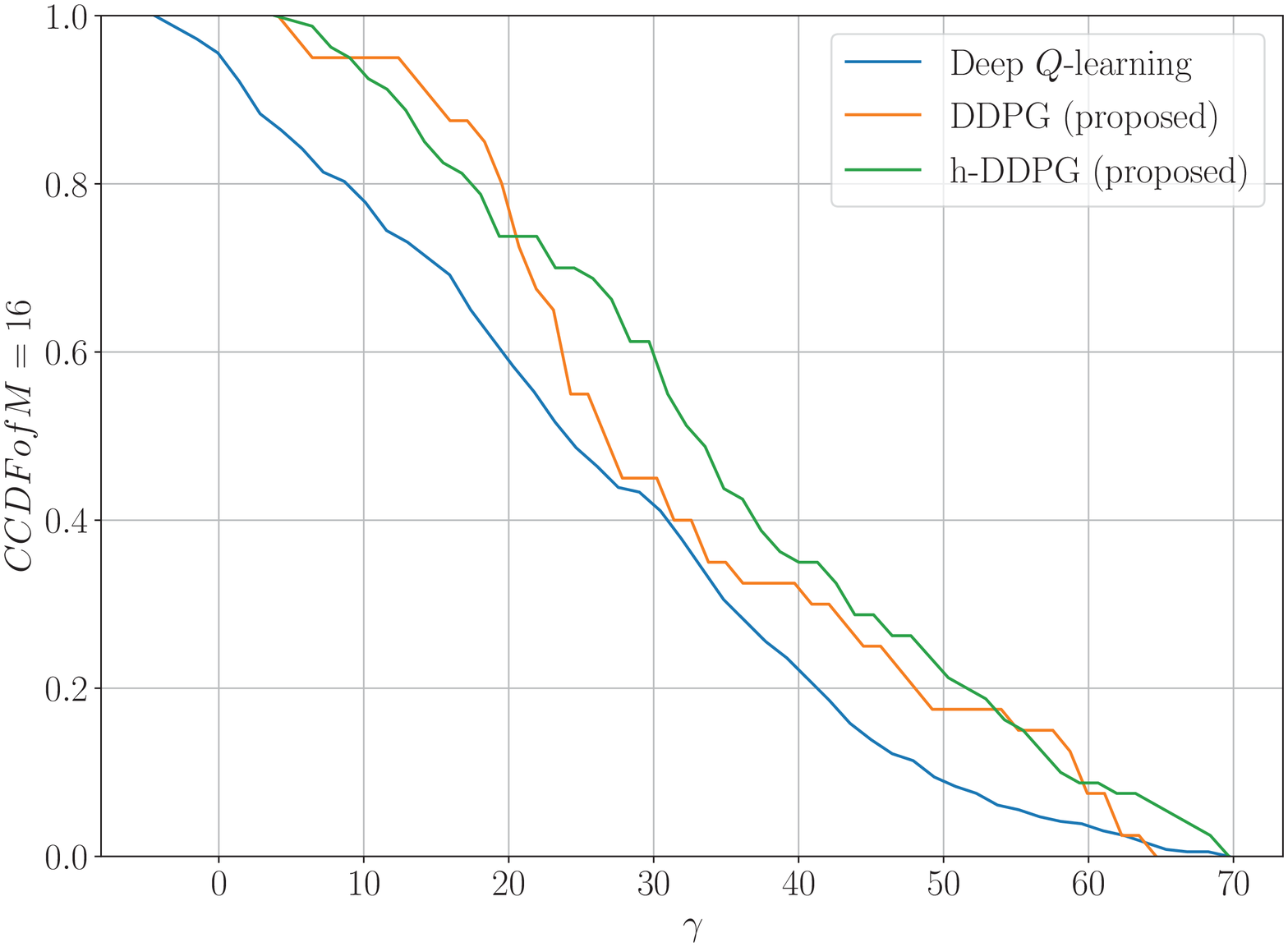}
}
\hfil
\subfloat[\textit{M}=32]{\includegraphics[width=2in,height=1.3in]{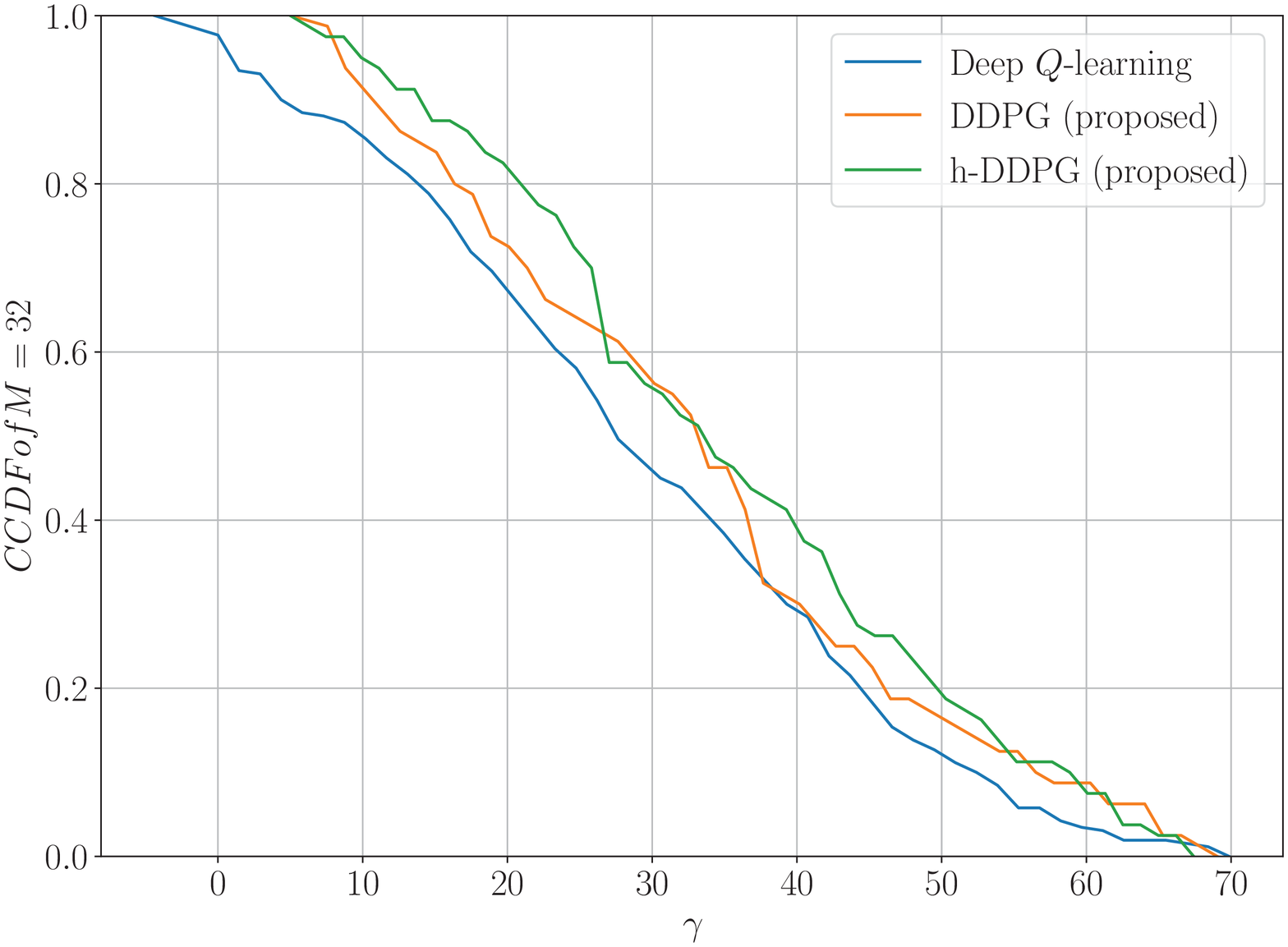}
}\hfil
\subfloat[\textit{M}=64]{\includegraphics[width=2in,height=1.3in]{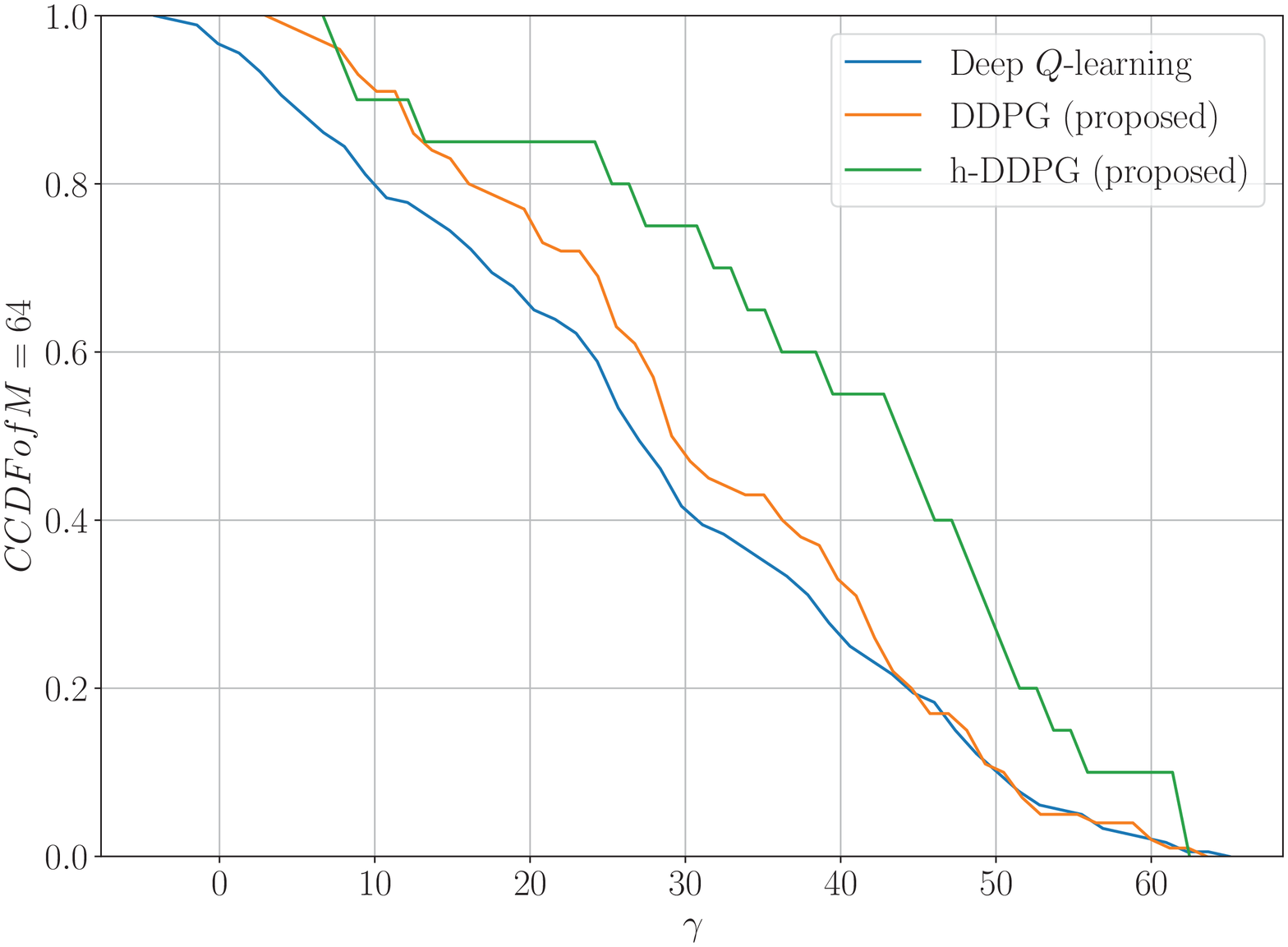}
}

\caption{The comparison of CCDF among DQN, DDPG and h-DDPG with the different number of antennas. (a)The CCDF of \textit{M}=4; (b)The CCDF of \textit{M}=8; (c)The CCDF of \textit{M}=16; (d)The CCDF of \textit{M}=32; (e)The CCDF of \textit{M}=64.}
\label{Fig.8}
\end{figure*}
\begin{figure*}[!t]
\centering
\subfloat[DQN]{\includegraphics[width=2in,height=1.3in]{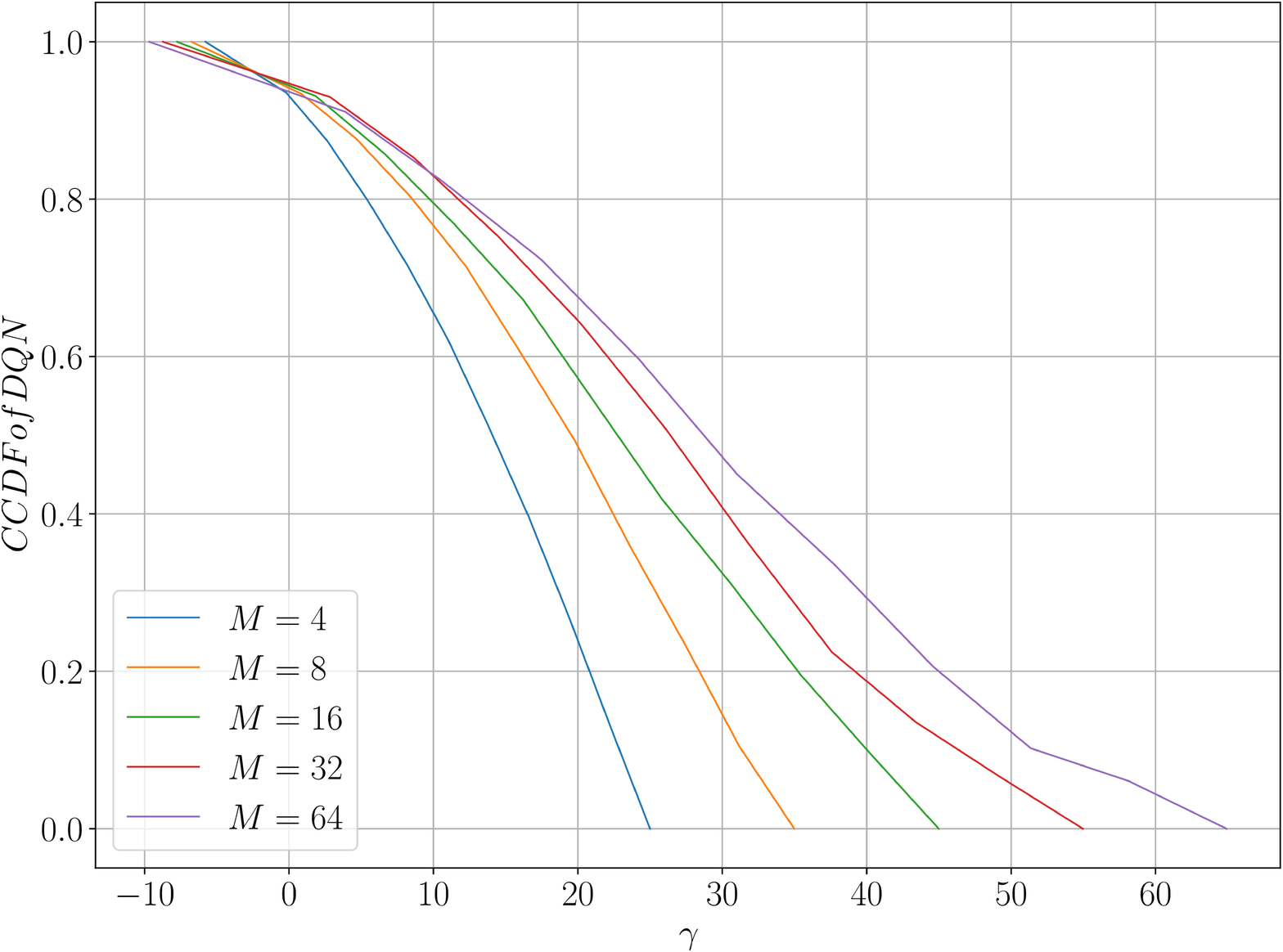}
}
\hfil
\subfloat[DDPG]{\includegraphics[width=2in,height=1.3in]{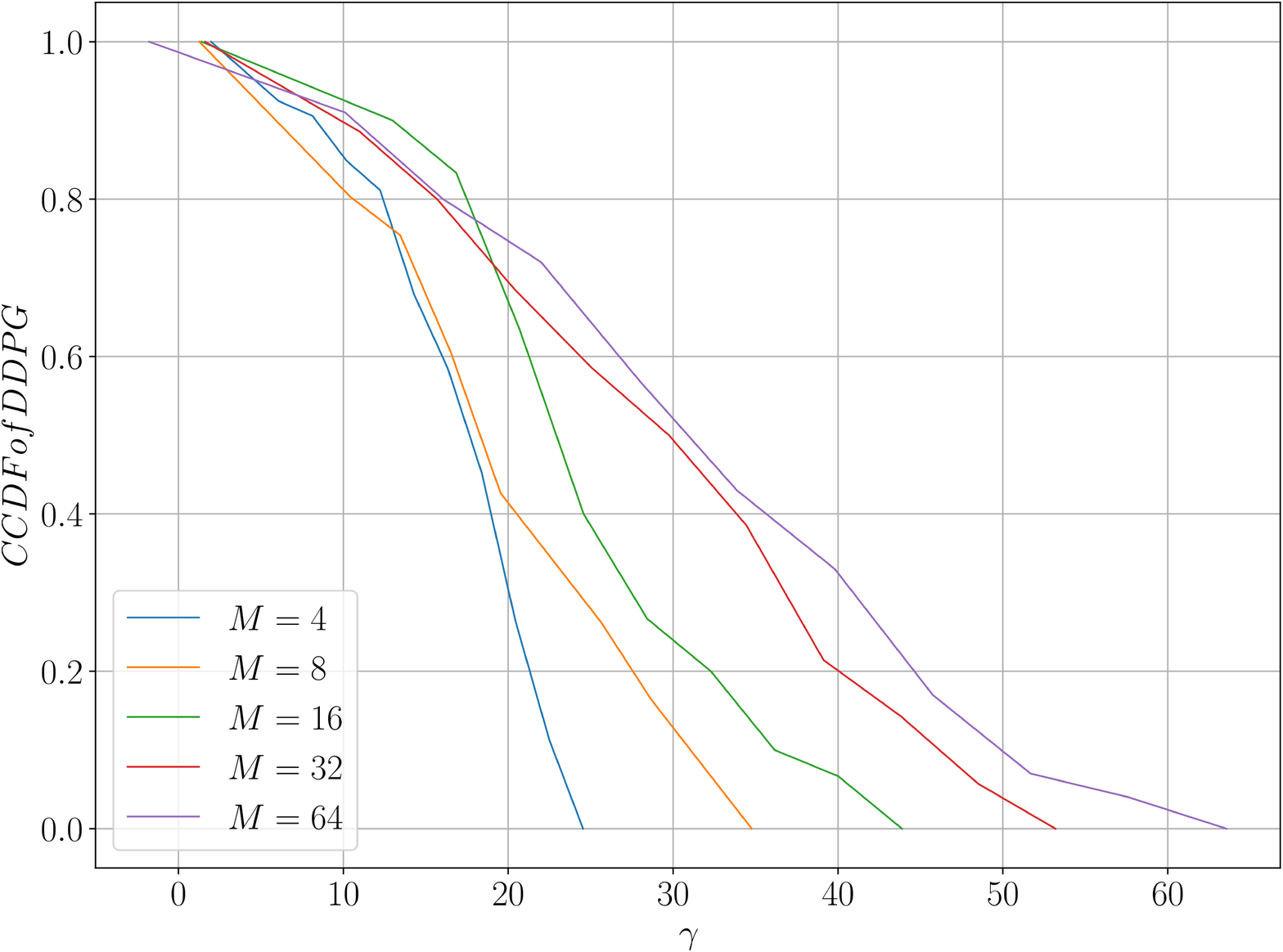}
}
\hfil\hfil
\subfloat[h-DDPG]{\includegraphics[width=2in,height=1.3in]{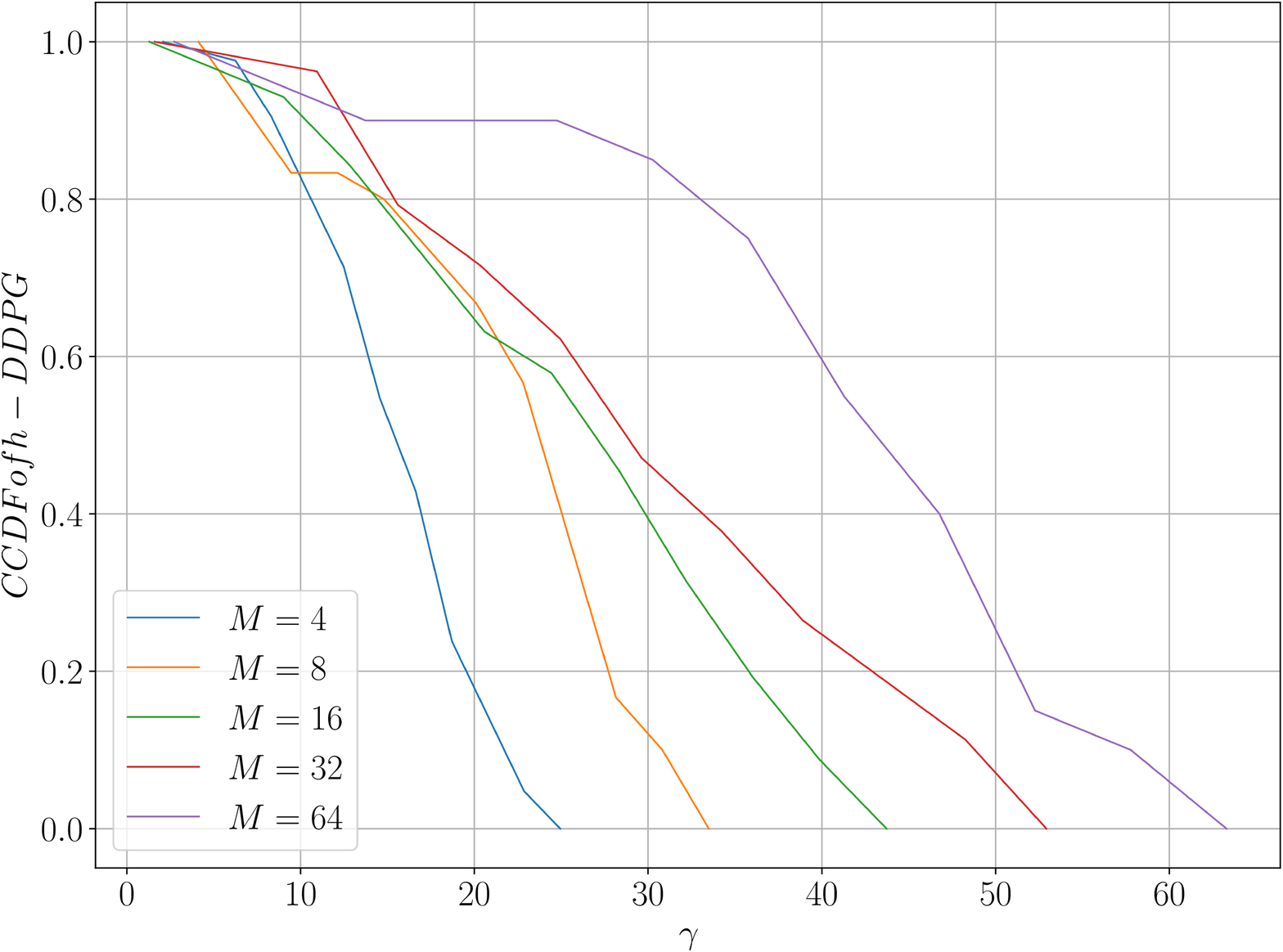}
}
\caption{ The comparison of CCDF among \textit{M}=4, \textit{M}=8, \textit{M}=16, \textit{M}=32 and \textit{M}=64 in different algorithms. (a)The CCDF of DQN; (b)The CCDF of DDPG; (c)The CCDF of h-DDPG.}
\label{Fig.9}
\end{figure*}

\begin{figure}[!t]
\centering 
\includegraphics[width=3in,height=2in]{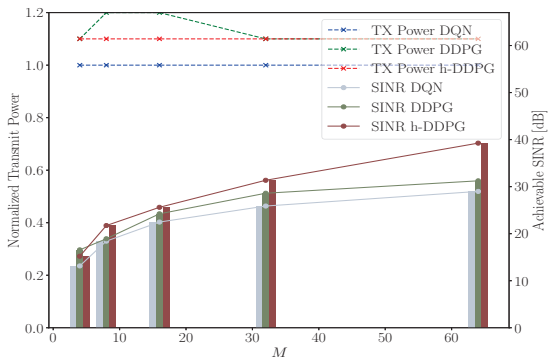}
\caption{Achievable SINR and normalized transmit power for DQN, DDPG and h-DDPG as a function of the number of antennas \textit{M}.}
\label{Fig.10}
\end{figure}

\begin{figure}[!t]
\centering 
\includegraphics[width=3in,height=2in]{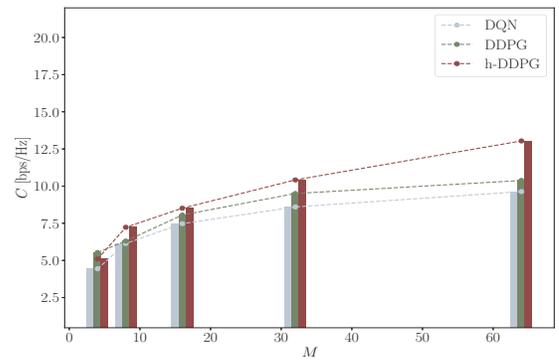}
\caption{Sum-rate for DQN, DDPG and h-DDPG as a function of the number of antennas \textit{M}.}
\label{Fig.11}
\end{figure}

\begin{figure*}[!t]
\centering
\subfloat[DQN]{\includegraphics[width=2in,height=1.3in]{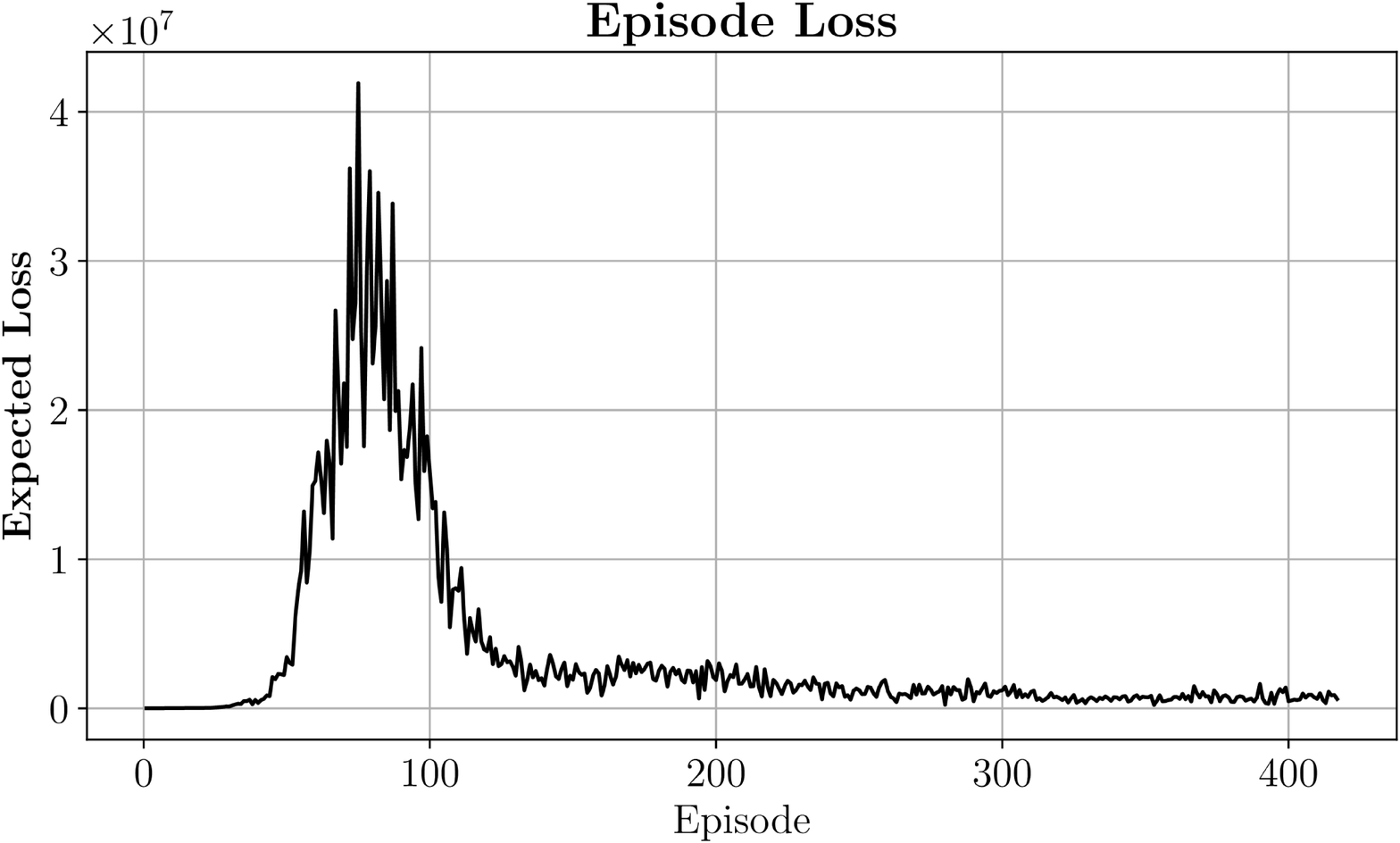}
}
\hfil
\subfloat[DDPG]{\includegraphics[width=2in,height=1.3in]{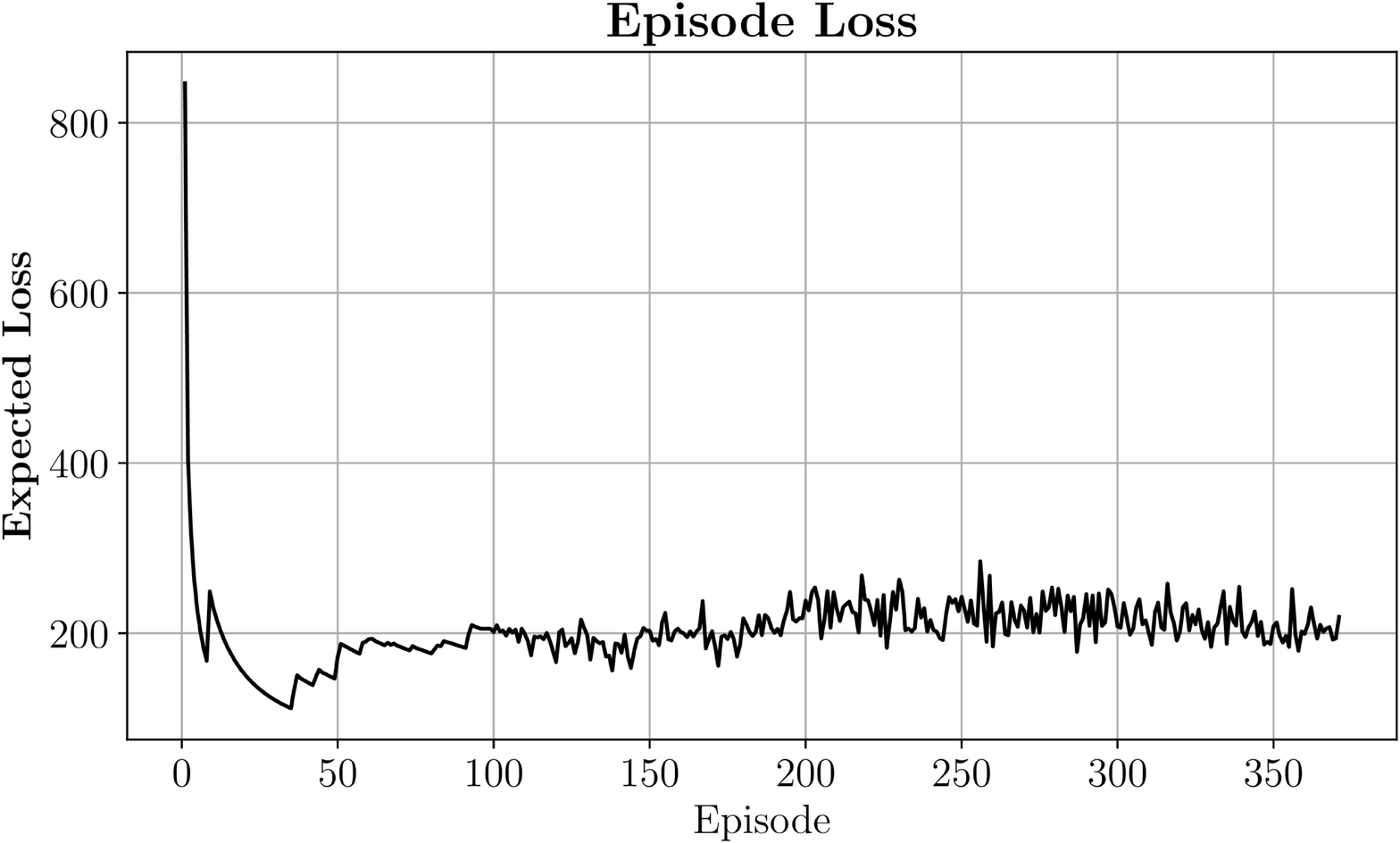}
}
\hfil\hfil
\subfloat[h-DDPG]{\includegraphics[width=2in,height=1.3in]{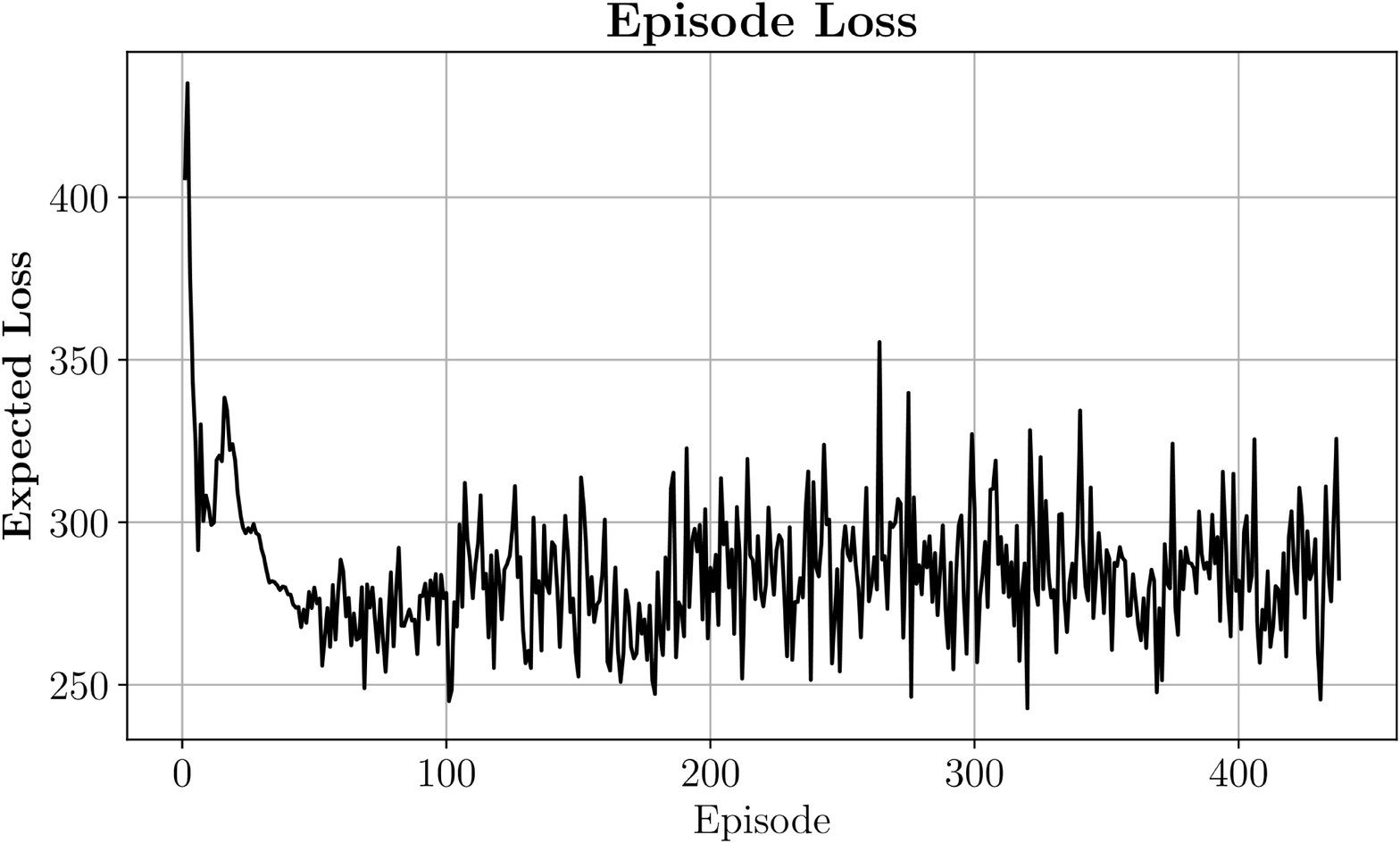}
}
\caption{The episode loss diagram of different algorithms. (a)The episode loss of DQN; (b)The episode loss of DDPG; (c)The episode loss of h-DDPG.}
\label{Fig.12}
\end{figure*}

\par Fig. \ref{Fig.7} shows the complement cumulative distribution function of $\gamma _{\rm{eff}}^l$ as the results from several algorithms in the simplest situation. It means there is just single antenna at each base station to avoid the discrete choice of the beamforming code book index and represent the improvement of proposed algorithms more intuitively. Obviously, the worst performance is from the FPA algorithm especially at the cell edge since FPA has no interference coordination but gives the dedicated transmit power of base station. The curve of Q-learning performs better than FPA and at least the possibility of $\gamma _{\rm{eff}}^l>$11dB is not under 0, which is because the PC and IC as effective measures are introduced to the model. Due to the limitation of the learning ability of Q-learning, the performance at $\gamma _{\rm{eff}}^l>$9dB is not excellent while DQN improves the coverage of received SINR by a wide margin since the use of neural networks in DQN increases the study ability of the model. What’s more, the implementation of the proposed algorithm one, DDPG, makes a great performance improvement based on DQN. This is because Q-learning and DQN just can adapt the transmit power of base stations with discrete increase or decrease so that the PC and IC cannot give the best performance in the model. Here the permission of DDPG for continuous actions solves this problem to some degree. Furthermore, the combination of hierarchical theory and DDPG has almost the best curve comparing with all other algorithms in our paper because the sparse reward problem due to the abortion of episodes are mitigated by the application of meta-policy. And it is not always for the model to meet this problem, which explains why hierarchical DDPG underperforms at $\gamma _{\rm{eff}}^l$=7dB.

When it comes to the complex model including beamforming, the superiority of DDPG and hierarchical DDPG is not very obvious since one of actions, beamforming, is discrete as the index of beamforming vector codebook, which reduces the improvement from the continuous action outputs of DDPG and hierarchical DDPG. As shown in Fig. \ref{Fig.8}, DDPG is generally better than DQN no matter how many antennas there are. And the performance of hierarchical is better than DDPG in most cases.

Looking form the other side, the CCDF, on behalf of the coverage, grows with the number \textit{M} of antennas for each algorithm for data bearers in Fig. \ref{Fig.9}. This is because the gain of beamforming array increases with \textit{M}, which is a function with \textit{M} as introduced before.

As shown in Fig. \ref{Fig.10}, the achievable SINR represents the performance directly, which also increases with \textit{M} since the formulation of beamforming array gain is ${\rm{|}}{f_a}|{|^2} \le M$ for each algorithm. What’s more, we can intuitively feel the improvement of received SINR by DDPG comparing with DQN and hierarchical DDPG increases the whole achievable SINR based on DDPG. The performance gap between DDPG and h-DDPG grows with \textit{M} because the range of index of beamforming codebook is much larger and the application of complex beamforming vectors at the base stations can lead to more abortion of episodes, which is the reason for the sparse reward problem. As for the transmit power, Fig. \ref{Fig.10} represents that the average value of transmit powers from two proposed algorithms are higher than DQN.

Fig. \ref{Fig.11} shows the sum-rate of these three algorithms. In general, DQN underperforms most while hierarchical DDPG is better than DDPG. And the performance gap between DDPG and h-DDPG grows with \textit{M} with the same reason as achievable SINR. Also, the sum-rate increases with \textit{M} for no matter which algorithm.

As shown in Fig. \ref{Fig.12}, the model of DDPG converges the fastest and hierarchical DDPG follows it. This is because the different goals made by meta controller influence the convergence of controller to some degree. As for the DQN, the reason is complex. The exploration of DQN takes advantage of $\varepsilon $-policy to carry out at random but not the deterministic policy from the neural network as Actor like DDPG. At the beginning, the samples are not enough for DQN and overfitting may occur in the model. In addition, DQN pursues the maximum of the state-action value but many state-action pairs never appear in the real situations, which would be corrected by the following learning. Therefore, the loss of DQN increases early and then decreases later.
\par Comparing the complexity of all algorithms, we find that the minimum is the one of FPA, just $O\left( {\rm{1}} \right)$, while the maximum is the complexity of Q-learning, which is $O\left( {{\rm{|}}S{\rm{||}}A{\rm{|}}} \right)$ due to the table of state-action values. As for DQN, DDPG and hierarchical DDPG, the complexity depends on the neural networks which is decided by the number of samples. Therefore, the overhead is $O\left( {T{N_{BS}}{N_{UE}}} \right)$ where ${N_{UE}}$ represents the total number of user equipment, ${N_{UE}}$ denotes the amount of base stations and \textit{T} is the periodicity of the measurements from any UE during time step $t$.

\section{Conclusions}\label{conclusions}
In this paper, we proposed two deep reinforcement learning methods, DDPG and hierarchical DDPG, to maximize the sum-rate for the time-varying downlink interference channel. On the one hand, the communication model with random routes is closer to the practical situation comparing the time-invariant channel or time-varying channel with fixed routes. On the other hand, the application of DDPG and hierarchical DDPG improves the performance and generalization of the model from multiple perspectives. Concretely, DDPG allows the continuous output by adapting the concept of Deterministic Policy Gradient while hierarchical DDPG solves the sparse reward problem by introducing the hierarchical theory.
\par In the simulation, industrial standard method Fixed Power Allocation, traditional reinforcement learning Q-learning, Deep Q-learning, our first proposed algorithm DDPG and the second proposed algorithm hierarchical DDPG respectively trained the simplest model. Here hierarchical DDPG performs best followed by DDPG, and DQN underperforms DDPG but is much better than Q-learning and FPA.
\par When it comes to the complex situation with joint power control, interference coordination and beamforming, we applied DQN, DDPG and hierarchical DDPG respectively. As for the convergence of the learning model, DDPG performs the best while hierarchical DDPG is better than DQN. And for the achievable SINR and sum-rate, DQN is the worst while hierarchical DDPG performs better than DDPG. In addition, the performance gap between hierarchical DDPG and DDPG grows with the number \textit{M} of antennas due to the relationship between the complexity of beamforming and \textit{M}. As for the coverage, hierarchical DDPG is the best while DDPG performs better than DQN.

\bibliographystyle{IEEEtran}
\bibliography{refer}

\end{document}